\documentclass[onecolumn,authoryear]{els-mrw} 

\usepackage{amsmath,amssymb,amsfonts,amsthm,makeidx,graphicx}
\usepackage{txfonts}
\usepackage{helvet}

\newcommand*\aap{A\&A}

\newcommand*\aapr{A\&A~Rev.}

\newcommand*\aj{AJ}

\newcommand*\apj{ApJ}
\newcommand*\apjl{ApJ}

\newcommand*\apjs{ApJS}

\newcommand*\apss{Ap\&SS}
\newcommand*\araa{ARA\&A}

\newcommand*\mnras{MNRAS}

\newcommand*\nat{Nature}

\newcommand*\pasp{PASP}

\newcommand*\rmxaa{Rev. Mexicana Astron. Astrofis.}

\begin{document}

\chapter{The Chemical Evolution of Galaxies}\label{chap1}

\author[1]{Mirko Curti}%


\address[1]{\orgname{European Southern Observatory}, \orgaddress{Karl-Schwarzschild-Strasse 2, 85748 Garching, Germany}}


\maketitle




\begin{glossary}[Nomenclature]
\begin{tabular}{@{}lp{34pc}@{}}
4MOST & 4-metre Multi-Object Spectroscopic Telescope\\
ADF & Abundance Discrepancy Factor \\
AGN & Active galactic nucleus \\
APOGEE & APO Galactic Evolution Experiment \\
BPT & Baldwin–Phillips–Terlevich \\
CC-SNe & Core-collapse supernovae \\
CEL & Collisionally-excited lines \\
CEMP & Carbon-enhanced metal-poor stars \\
CGM & Circumgalactic medium \\
DLA & Damped Lyman-$\alpha$ system \\
E-ELT & European extremely large telescope \\
ESO & European Southern Observatory \\
EW & Equivalent width \\
FMR & Fundamental metallicity relation \\
ICM & Intracluster medium \\
IFU & Integral field unit \\
IFS & Integral field spectroscopy \\
IGM & Intergalactic medium \\
IMF & Initial mass function \\
IR & Infrared \\
ISM & Interstellar medium \\
JWST & James Webb Space Telescope \\
LLS & Lyman Limit System \\
MaNGA & Mapping nearby galaxies at APO \\
MOONS & Multi object optical and near-infrared spectrograph \\
MUSE & Multi unit spectroscopic explorer \\
MW & Milky Way \\
MZR & Mass–metallicity relation \\
RL & Recombination lines \\
SAMI & Sydney-AAO Multi-object Integral field spectrograph \\
SDSS & Sloan digital sky survey \\
SFR & Star formation rate \\
SSP & Single stellar population \\
UV & Ultraviolet \\
VLT & Very large telescope \\

\end{tabular}
\end{glossary}

\begin{abstract}[Abstract]
Metals—heavy elements synthesized during various phases of stellar evolution or during supernova explosions—play a fundamental role in shaping galaxy evolution.
In fact, their relative abundances, spatial distribution, and scaling with galactic properties reflect the constant interplay between star-formation, nucleosynthesis, and gas flows that drive the cycle of baryons in-and-out of galaxies across the cosmic time.
This chapter aims at offering a concise introduction to the methodologies used to measure elemental abundances in galaxies and the basics of chemical evolution modeling. 
We also provide a brief overview of the current observational framework, including metallicity scaling relations, the study of relative chemical abundances, the distribution of metals within and beyond galaxies, and how these properties evolve with redshift, 
drawing on both extensive literature and recent developments, and aiming to highlight well-established findings alongside ongoing challenges in this rapidly advancing field.
\end{abstract}

\begin{glossary}[Keywords]

Galaxies:Emission line galaxies; Galaxies:High-redshift galaxies; Galaxy physics:Galaxy formation; Galaxy physics: Galaxy abundances; Chemical Abundances:Metallicity; Chemical Abundances:Abundance ratios

 \end{glossary}

\begin{BoxTypeA}[chap1:box1]{Key points}
\begin{itemize}
    \item \textbf{Metallicity} The abundance of all elements heavier than hydrogen and helium relative to the total mass of baryons. Often expressed in terms of relative number densities.
    \item \textbf{Oxygen abundance} The relative abundance of oxygen compared to hydrogen. It is often adopted as a proxy for the total metallicity of a galaxy.
     \item \textbf{Stellar metalliicty} The mass-weighted or luminosity-weighted metallicity of stellar populations. It reflects the integrated star-formation history of a galaxy.
    \item \textbf{Gas-phase metallicity} The metallicity of the gas-phase of the ISM, reflecting the `current' level of chemical enrichment and probing recent star-formation episodes and short-timescales processes and gas flows.
    \item \textbf{Te-method} A methodology to derive gas-phase abundances from the direct measurement of the temperature of free electrons in the ionized gas
    \item \textbf{Strong-line methods} Ratios of bright emission lines in galaxy spectra calibrated as gas-phase metallicity diagnostics
    \item \textbf{Absorption line system} A gas cloud observed against a background source of radiation, producing absorption lines in the spectrum whose properties depend on the intervening column density
    \item \textbf{Damped Lyman alpha} Absorption system with column density $\mathrm{N_{HI} > 10^{20.3} cm^{-2}}$ 
    \item \textbf{Closed-box model} Analytical chemical evolution model assuming no gas flows in-and-out of the system
    \item \textbf{Stellar yield}  The amount of metals
    produced by a stellar generation and returned to the ISM 
    \item \textbf{Effective yield} The effective (lower) value of the stellar yield that account for metals expelled within gas outflows
    \item \textbf{Mass-Metallicity relation} The scaling relation between the stellar mass content and the metallicity (either stellar or gas-phase) observed in galaxy populations.
    \item \textbf{Fundamental metallicity relation} The scaling relation between stellar mass, metallicity, and star-formation rate. Galaxies with high SFR have on average lower metallicity than low-SFR systems for the same stellar mass. 
    \item \textbf{Chemical abundance patterns} Relative abundances of different elements. Their trends with total metallicity and cosmic time constrain the history of star-formation in galaxies.
    \item \textbf{Metallicity gradient} Trend of metal enrichment with galactocentric radius.
    \item \textbf{Metal budget} Accounting for the total amount of metals produced by galaxies by studying their distribution over different phases.
        
\end{itemize}
\end{BoxTypeA}

\section{Introduction}
\label{sec:intro}
In the $\Lambda$CDM cosmological framework, the Universe consists primarily of baryons, dark matter, and dark energy, which collectively influence the shape of space and the arrangement of cosmic structures. Structure formation is driven by gravity, which amplifies small irregularities in the dark matter’s otherwise uniform distribution. When these irregularities grow to a certain size, they form dark matter halos—gravitationally bound regions that continue to expand through matter accumulation and mergers with smaller structures. As gas cools within these halos, conditions emerge that enable star formation, leading to the first massive, metal-free (Population III) stars. These early stars quickly exhausted their fuel, producing the first metals that enriched the surrounding interstellar medium (ISM). 

Once they have formed, galaxies continuously undergo chemical enrichment. Metals produced via stellar nucleosynthesis or during violent Supernovae explosions are then returned to the ISM, enriching the galactic environment. In addition, cosmological accretion from the cosmic web and outflows driven by either star-formation or super-massive black holes activity further regulate the metal content of the ISM. 
The continuous interplay among these processes drives the baryonic cycle in and out of galaxies, and shapes the scaling relations between metallicity and other global properties which are observed in galaxy populations in both the nearby Universe and at early cosmic epochs.
Furthermore, different chemical species are produced (and returned to the ISM) on different timescales by stars of different masses; therefore, the relative abundance of elements provides unique and complementary insights on the star-formation history of galaxies.
Overall, the analysis of the chemical properties of galaxies and their evolution with cosmic time deliver therefore key observational constraints that need to be matched by any model and simulation aimed at describing how galaxies form and evolve.

Throughout this Chapter, we will provide an overview on the chemical properties of galaxies in both the local and early Universe.
We will first discuss the main techniques adopted to measure chemical abundances in stellar populations and in the interstellar medium of galaxies, addressing strengths and weaknesses of the different methods.
Then, we will introduce some simple realizations of chemical evolution models, and discuss how the parameters in these models are directly linked to the physical processes driving the mass assembly, star-formation, and chemical enrichment in galaxies over cosmic time.
Finally, we will present a brief overview of the current observational landscape of the field (including some of the most recent developments from the James Webb Space Telescope, JWST), focusing in particular on the metallicity scaling relations in galaxies and their evolution with redshift, the study of different chemical abundance patterns, and the distribution of metals inside and outside galaxies.

\section{Metallicity measurements in galaxies}
\label{sec:met_measurements}

\subsection{Some definitions}
\label{ssec:met_definition}
In the astrophysical context, the word `metals' refers to all chemical elements heavier than hydrogen and helium.
By strict definition, the `metallicity' indicates the abundance by mass of all metals relative to the total mass of baryons (which is dominated by hydrogen and helium).
However, in galaxy evolution studies the relative abundance of two elements is generally expressed in terms of their relative number densities, such that 
\begin{align}
    \mathrm{X/Y} \equiv \mathrm{N_X / N_Y}.
\end{align}
When expressing the relative abundance of a given ionic species relative to hydrogen, the following notation is also commonly adopted
\begin{align}
    \mathrm{12+log(X/H)} \equiv \mathrm{12+log(N_X / N_H)} \, 
\end{align}
where the value $12$ is added to ensure such expression provides always a positive value, or
\begin{align}
    \mathrm{[X/H]} \equiv \mathrm{log(N_X / N_H)} - \mathrm{log(N_X / N_H)}_{\odot} \, 
\end{align}
when expressing quantities relative to Solar abundances. 
Since measuring elemental abundances is often observationally challenging, in galaxy evolution studies it is common practice to assume the oxygen abundance ($\mathrm{12+log(O/H)}$) as a proxy of the total metallicity, as oxygen is generally the most abundant heavy element by mass and it is responsible for the brightest emission lines seen in galaxy spectra (hence it is often the only element whose abundance can be reliably measured). This approximation assumes that the abundances of all other heavy elements scale proportionally to Solar abundance ratios which, as we will see later on, it might often be an incorrect assumption in early galaxies.
Furthermore, the exact value of the Solar oxygen abundance can vary depending on the measurement techniques, in that for instance hydrodynamical models of the solar atmosphere provides a photospheric oxygen abundance of $\mathrm{12+log(O/H)_{\odot} = 8.69\pm0.05}$ \citep{asplund_solar_2009}, whereas studies based on helioseismology delivers higher values, i.e. $\mathrm{12+log(O/H)_{\odot} = 8.86\pm0.05}$ \citep{delahaye_solar_2006}.
Nonetheless, in the framework of galaxy evolution studies the absolute Solar abundance values should only be considered as a reference, although care should be taken in rescaling observational results of chemical abundances in galaxies to the same Solar abundance scale when comparing different studies.

\subsection{Stellar metallicity}
\label{ssec:stellar_met}
The metallicity of the stellar populations provides information on the integrated star-formation history of galaxies, as it reflects the total amount of heavy elements built throughout subsequent stellar generations across galaxy lifetimes and currently locked into stars.
Elemental abundances in the stellar photospheres are reflected into a wealth of characteristic absorption features observable in the continuum of galaxy spectra, and which have been extensively calibrated against models to identify robust stellar metallicity diagnostics.   

\subsubsection{Spectral absorption features and indices}
\label{sssec:stellar_indices}

Historically, early approaches involved the development of a set of spectral indices, including the so-called Lick indices, defined as the ratio of the flux measured within a given bandpass (tipically $\sim \mathrm{50\AA}$ width) and that of a nearby pseudo-continuum evaluated in two nearby wavelength windows, such as \citep[following the definition of]{Worthey_lick_indices_1994} : 
\begin{align}
    \text{Index} = \int_{\lambda_1}^{\lambda_2} \left(1 - \frac{f_\lambda}{F_C}\right) d\lambda \ ,
\end{align}
where
$\mathrm{f_\lambda}$ is the observed flux density at wavelength $\lambda$, $\lambda_1$ and $\lambda_2$ are the boundaries of the feature bandpass, and with the pseudo-continuum $\mathrm{F_C}$ estimated as
\begin{align}
F_C = \frac{\lambda_r - \lambda}{\lambda_r - \lambda_b} \cdot \left\langle f_\lambda \right\rangle_{\lambda_b} + \frac{\lambda - \lambda_b}{\lambda_r - \lambda_b} \cdot \left\langle f_\lambda \right\rangle_{\lambda_r} \ ,
\end{align}
where
$\left\langle f_\lambda \right\rangle_{\lambda_b}$ is the average flux density in the blue pseudo-continuum bandpass,  
$\left\langle f_\lambda \right\rangle_{\lambda_r}$ is the average flux density in the red pseudo-continuum bandpass, 
and $\lambda_b$, $\lambda_r$ are the central wavelengths of the blue  and red bandpasses, respectively. 

In the rest-frame optical spectrum, widely used indices include features sensitive to both the age of stellar populations (such as the depth of the break at $\mathrm{4000}\AA$, D$_{n}$(4000) and of the Balmer absorption features such as $\mathrm{H\gamma_{A}}$ and $\mathrm{H\beta}$), as well as features sensitive to the abundances of specific elements like magnesium (e.g., Mg1, Mg2, Mgb), calcium (e.g., the calcium triplet, CaT), and iron (e.g., $\mathrm{Fe5270}$, Fe5335) which are calibrated to infer the metallicity of unresolved stellar populations in nearby galaxies.

A wealth of information about the metallicity of young, massive stars is also encompassed by features in the rest-frame UV region of the spectrum.
In fact, they do not only reflect the composition of the photospheres of O and B stars, but also stellar winds whose strength increase with increasing metallicity due to metals dominating the opacity of the stellar material, hence allowing radiation pressure to more effectively drive material off the surface of the star. 
Among the most commonly analysed metallicity-sensitive, rest-frame UV stellar features are stellar-wind lines such as $\mathrm{NV\lambda 1238,1242}$, SiIV$\lambda$1393,1402, CIV$\lambda$1548,1550, and HeII$\lambda$1640, or photospheric lines such as FeV$\lambda$1360-1380, OV$\lambda$1371, SiIII$\lambda$1417, CIII$\lambda$1427, FeV$\lambda$1430, SiII$\lambda$1533, and CIII$\lambda$2300 \citep{leitherer_2011}.
However, these features depend not only on the metallicity of stellar population, but also on factors such as stellar age and the shape of the initial mass function (IMF). 
Indeed, and similar to what done for rest-optical features, several spectral indices have been identified in the attempt to better isolate the metallicity dependence, by either leveraging empirical spectral libraries or sets of theoretical stellar spectra \citep[e.g.][]{Byler_diagnostics_2018}.

\subsubsection{Full spectral fitting techniques}
\label{sssec:full_sed}

Over the past two decades, there has been extensive development of tools aimed at inferring multiple physical properties by reproducing the full Spectral Energy Distribution (SED) of galaxies, from ultraviolet to infrared wavelengths \citep[see e.g.][]{conroy_modeling_2013}.
These tools generally employ either empirical or theoretical libraries of stellar templates sensitive to different physical parameters (e.g., age, IMF, metallicity, dust, star-formation rate), which are then combined to generate a model SED which is compared to data.
The model SED can either represent a `simple stellar population' (SSP),  describing the evolution in time of the SED of a single, coeval stellar population of a given metallicity, or combine multiple SSPs into more complex `composite stellar populations' (CSPs), which contain stars with a range of ages and metallicities (as well as the contribution from dust) that reflect the star-formation history (SFH) of the galaxy.
Both empirical and theoretical libraries come with their own intrinsic strengths and weaknesses, and there is no single library that covers the entire range of parameter space necessary for constructing SPS models, often requiring to combine together various libraries of different quality.
Theoretical libraries offer the great advantage of densely covering the parameter space, they are not limited in spectral resolution, and their model spectra are not subject to typical flux calibration issues. The  disadvantage is that such libraries are only as good as the input atomic and molecular parameters, and are strongly dependent on the approximations made in the treatment of different physical processes (e.g. convection and  microturbulence), as well as departures from local thermodynamic equilibrium (LTE) and from simple, plane-parallel geometry. On the other hand, empirical spectra are affected by observational constraints, being limited in wavelength coverage and spectral resolution and often largely incomplete in their probe of the physical parameter space. 

These stellar population synthesis models are then fit to observational data (either broadband SEDs or spectra) to constrain multiple physical properties, including metallicity, dust attenuation, and parameters used to model the galaxy SFH. The fitting approach generally employs standard $\mathrm{\chi^2}$-minimisation techniques \citep[e.g. \textsc{ppxf}, ][]{cappellari_improving_2017}, or is performed within full Bayesian frameworks in which the parameter space is sampled exploiting algorithms such as Markov Chain Monte Carlo (MCMC) and the best-fit values are inferred from the marginalized posterior distributions for each parameter \footnote{however, care should always be taken in considering the (sometimes significant) influence that the specific choice of the priors can have on the inferred best-fit values} (e.g. \textsc{beagle}, \citealt{chevallard_beagle_2016}, \textsc{bagpipes}, \citealt{carnall_bagpipes_2018}, \textsc{cigale}, \citealt{Boquien_cigale_2019}, \textsc{prospector}, \citealt{Johnson_prospector_2021}).

\subsection{Metallicity of the gas-phase}
\label{ssec:gas_met}

The metallicity of the gas-phase provides an ‘instantaneous’ snapshot of the recent SFH in galaxies, as it does not only reflect the enrichment of the gas around recently formed O and B stars, but it is also sensitive to recent gas inflows and outflows episodes.
Generally, the ISM metallicity is mostly probed by its ionised phase, because the gas illuminated by young stars in H$\mathrm{II}$ regions produces a wealth of emission lines which are easily observable in galaxy spectra, and from which the elemental abundance of different ionic species (in particular oxygen, as discussed in Sect.~\ref{ssec:met_definition}) can be derived by means of different techniques, as we will briefly describe in this Section.

\subsubsection{The electron-temperature method}
\label{sssec:te_method}

At the base of this method lies the principle that the flux of a given emission line as observed in a galaxy spectrum is the product of the abundance of the ionic species of that given element times its volumetric emissivity.
Therefore, the abundance ratio of two ions can be obtained from the ratio of the observed line intensities, once the different temperature- and density-dependence of the emissivities can be properly accounted for : 
\begin{align}
    \frac{\mathrm{N(X^l)}}{\mathrm{N(Y^m)}} = \frac{\mathrm{I_{\lambda,X^l}}}{\mathrm{I_{\lambda,Y^m}}} \cdot \frac{\mathrm{J_{\lambda, Y^m} (T_e,n_e)}}{\mathrm{J_{\lambda,X^l} (T_e,n_e)}} \ ,
\end{align}
where $\mathrm{I_{\lambda,X^l}}$ is the intensity of an emission line at wavelength $\lambda$ for the ionic species $\mathrm{X^l}$, and $\mathrm{J_{\lambda,X^l}}$ is the emissivity.
When deriving the relative abundances of specific metal ionic species (for instance, doubly-ionized oxygen, $\mathrm{O^{++}}$) to hydrogen, the above formula becomes 
\begin{align}
   \frac{\mathrm{O^{++}}}{\mathrm{H^+}} = \frac{\mathrm{I([\text{O III}] \lambda 5007)}}{\mathrm{I(H\beta)}} \cdot \frac{\mathrm{J_{H\beta}(T_e, n_e)}}{\mathrm{J_{[\text{O III}]}(T_e, n_e)}} \ ,
\end{align}
where it is common practice to measure line strengths relative to $\mathrm{H\beta}$ because this line is readily observed and generally not blended with other lines even at lower spectral resolutions.
Most emission lines from metal ionic species at optical wavelengths are forbidden lines arising from collisional excitation (CEL), with critical densities\footnote{defined as the density at which the rate of collisional de-excitations equals the rate of spontaneous radiative transitions} generally much higher than the density of medium they originate from. In such circumstances, the volumetric emissivity of the transition is given by
\begin{align}
    \mathrm{J_{\lambda} = h\nu_{ul} \frac{n_u A_{ul}}{4\pi} \sim n_e n_x e^{-E_{ul}/kT_e}} \ ,
\end{align}
where $\mathrm{n_e}$ is the density of electrons (i.e. the `colliding particles'), $\mathrm{n_x}$ the density of the ion X, $u$ and $l$ the upper and lower levels of the transition, and A the Einstein coefficient for spontaneous emission, hence showing that $\mathrm{J_{\lambda}}$ has an exponential dependence on temperature.
On the contrary, the temperature-dependence of the volumetric emissivity of a permitted, recombination line such as $\mathrm{H\beta}$ is only (approximately) linear.
Therefore, when comparing the intensity of a CEL to that of a recombination line to infer ionic abundances, one needs to account for the fact that the ratio of the volumetric emissivities is a strong function of temperature (and, though more mildly, of density), hence the application of this method (often known as the `$\mathrm{T_e}$-method') requires accurate knowledge of both electron temperature $\mathrm{T_e}$ and density $\mathrm{n_e}$ of the emitting nebula.
The total elemental abundance is then obtained by co-adding the individual ionic abundances, and correcting for those not directly observed by means of specific `ionization correction factors' (ICF).

\begin{figure}
    \centering
    \includegraphics[width=0.95\linewidth]{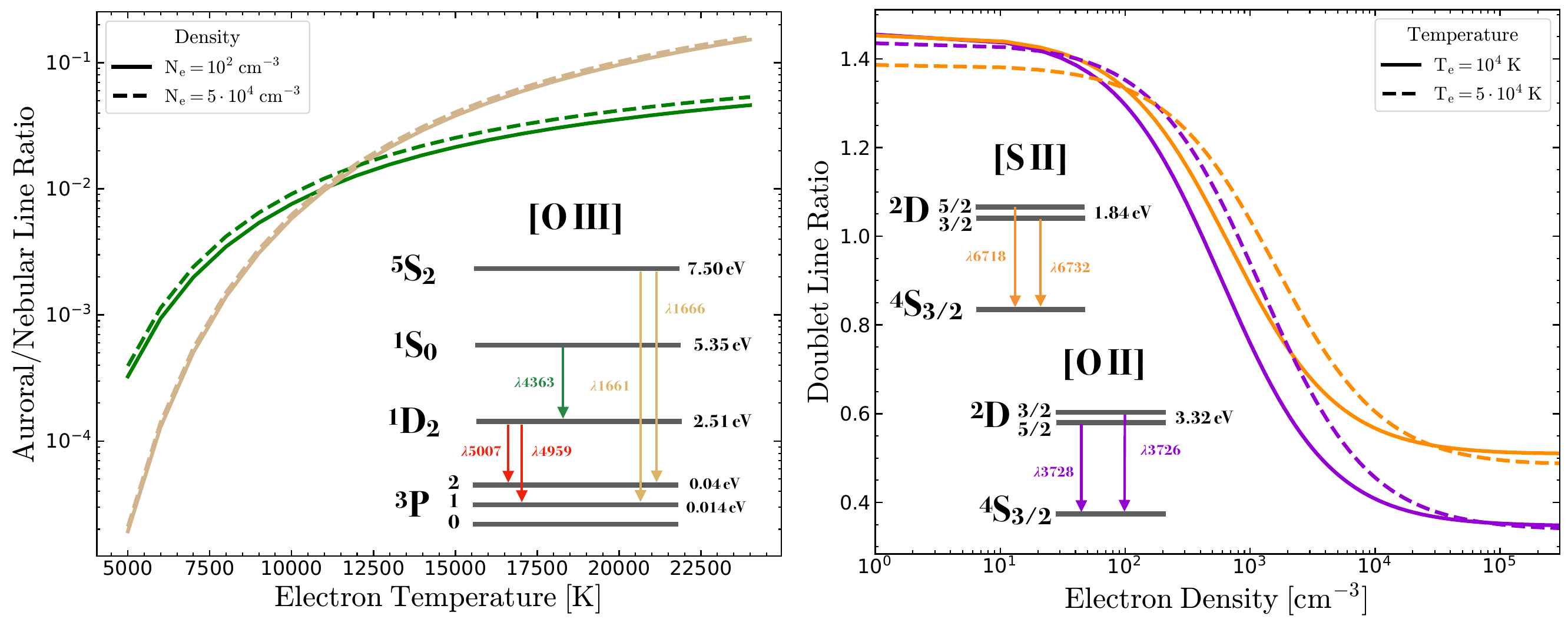}
    \caption{Examples of diagnostics of the ISM. The left-hand panel reports the behavior of different $\mathrm{[O III]}$ auroral-to-nebular line ratios as a function of temperature, namely $\mathrm{[O III]\lambda4363/[O III]\lambda5007}$ and $\mathrm{[O III]\lambda\lambda1661,1666/[O III]\lambda5007}$. Each curve is plotted for two very different electron density regimes, demonstrating the small residual dependence on $\mathrm{N_e}$. The Grotrian diagram for $\mathrm{[O III]}$, highlighting the involved transitions, is plotted in the bottom-right part of the panel. The right-hand panel shows instead two classical density diagnostics, namely the $\mathrm{[O II]\lambda3727/[O II]\lambda3729}$ and $\mathrm{[S II]\lambda6718/[O II]\lambda6732}$ doublet line ratios. The curves are plotted for two different $\mathrm{T_e}$ values, and the Grotrian diagrams of the involved transitions are also shown.
    }
    \label{fig:te_ne_diagnostics}
\end{figure}

\paragraph{\textbf{Temperature and density diagnostics}}
The structure of the energy levels of some ionic species is such that the ratio between different optical emission lines can be employed as temperature or density diagnostics.
For instance, the energy-levels configuration of the $\mathrm{O^{++}}$ ion is represented in the left panel of Fig.~\ref{fig:te_ne_diagnostics}. 
All the levels above the ground state are populated by the excitation due to the collisions between ions and free electrons. The relative populations of the $\mathrm{^1S}$ and $\mathrm{^1D}$ states depend critically on $\mathrm{T_e}$, hence the ratio between emission lines arising from these levels will be highly temperature-sensitive. For instance, in nebular and galaxy evolution studies the $\mathrm{[O III]\lambda 4363/[O III] \lambda 5007}$ ratio is one of the most commonly adopted diagnostics of electron temperature. We note that in this context lines originating from the highest energy levels (e.g. $\mathrm{[O III]\lambda 4363}$, or the rest-frame UV $\mathrm{O III]\lambda\lambda 1661,1666}$) are often referred to as `auroral lines'; such emission lines are generally much fainter than `nebular' transitions (e.g. $\mathrm{[O III]\lambda 5007}$), down to a factor of $\approx$~a few hundredths at high metallicity\footnote{gas temperature and metallicity are tightly correlated, as emission lines from metal ions are the primary cooling channels of the ISM}, hence often representing the biggest observational limitation to the application of the $\mathrm{T_e}$-method. 
Other common $\mathrm{T_e}$ diagnostics involving different ionic species (and probing different ionization states of the gas) include the  $\mathrm{[N II]\lambda 5755/[N II] \lambda 6584}$, the $\mathrm{[O II]\lambda\lambda 7320,7330/[O II] \lambda\lambda 3727,3729}$, and the $\mathrm{[S III]\lambda 6312/[S III] \lambda 9530}$ line ratios. From detailed calculations of (approximated) 5-level atom structures, numerical formulae to measure the electron temperature from the observed line ratios are derived; however, we note that these expressions  depends on the assumed collision strengths of the transitions, which are generally tabulated for the typical range of nebular conditions but can vary between different versions of atomic datasets. 

The electron density $\mathrm{n_e}$ can be measured instead exploiting ratios of emission lines originating from excited states of similar energy (hence emitted in wavelength-close doublets), but different critical densities for collisional de-excitation.
These ratios are sensitive to $\mathrm{n_e}$ (while almost insensitive to temperature) in the regime comprised between the critical densities of the two transitions, while saturate in both lower and higher density regimes.
Typical density diagnostics for H$\mathrm{II}$ regions in rest-frame optical are the $\mathrm{[O II]\lambda\lambda 3727,3729}$ and $\mathrm{[S II]\lambda\lambda 6717,6731}$ doublets (right-hand panel of Fig.~\ref{fig:te_ne_diagnostics}), while diagnostics in the rest-frame UV, sensitive to higher density regimes, include e.g. the $\mathrm{C III]\lambda\lambda 1907,1909}$ and $\mathrm{N IV]\lambda\lambda 1483,1485}$ doublets.

\paragraph{\textbf{Caveats on the $\mathrm{T_e}$-method}}
A number of assumptions underlines the application of the $\mathrm{T_e}$-method. Since forbidden, collisionally-excited lines are optically thin, any flux measurement represents emissivity-weighted averages along the line of sight through the nebula. Although in the case of reasonably homogeneous conditions the measured electron temperature is representative of the average temperature of the nebula, inhomogeneities in the form of temperature and density fluctuations can instead severely bias the interpretation of the inferred temperature. 
It is indeed well know that H$\mathrm{II}$ regions are far from being homogeneous: on the contrary, pockets, filaments, and shells with different $\mathrm{n_e}$ and  $\mathrm{T_e}$ structures are clearly observed \citep{mendez_delgado_DESIRED_HII_2023, Kreckel_LVM_orion_2024}. In regions where $\mathrm{T_e}$ is enhanced, the emissivity of the temperature-sensitive auroral lines is disproportionally enhanced over that of the stronger nebular lines: this translates into an overestimation of the average $\mathrm{T_e}$, leading to an underestimation of the metallicity when integrating the line ratios over the whole nebula.
Such effect is even more prominent when considering global, integrated galaxy spectra which are the result of averaging over multiple H$\mathrm{II}$ regions of different properties, and which make the interpretation of metallicity as measured in galaxy spectra not straightforward \citep{pilyugin_electron_2010}. The advent of integral-field-spectroscopy (providing spatially-resolved spectral information for both nearby and distant galaxies) marked great advancements in this regard.

\paragraph{\textbf{Abundances from recombination lines and the `Abundance Discrepancy Factor'}}

In principle, the most `direct' probe of elemental abundances would follow by comparing recombination lines (RL) of metal ions with those of hydrogen.
In fact, the similar dependence of the involved emissivities on both $\mathrm{n_e}$ and $\mathrm{T_e}$ would make the ratio between RLs of different ions only proportional to the relative abundances, while almost independent on the gas density and temperature conditions, hence more robust against the typical biases of the $\mathrm{T_e}$-method associated with temperature fluctuations \citep[e.g.][]{peimbert_densities_2014}.
However, this also means that RLs of metal species are extremely faint (of the order of $10^{-3}$-$10^{-4}$ times the intensity of Balmer lines, even for the most abundant elements like carbon and oxygen): therefore, the detection of metal RLs and their applicability in abundance measurements is practically limited only to bright and nearby H$\mathrm{II}$ regions, planetary nebulae, and supernova remnants in the Milky Way (MW) and the Local Group \citep[e.g.][]{garcia_rojas_esteban_2007}, and requires spectra of high-resolution and high signal-to-noise ratio. 

Notably, chemical abundances measured via RLs have been widely observed to deviate from those based on the $\mathrm{T_e}$-method, with typical discrepancies up to a factor of $\sim0.3$~dex. The origin of such disagreement, also known as `Abundance Discrepancy Factor' \citep[ADF, see e.g.][]{peimbert_nebulae_2017, toribio_san_cipriano_2017}, is still currently debated.
On the one hand, temperature fluctuations are invoked as the main culprits, affecting, as seen, mostly $\mathrm{T_e}$ measurements, and inducing an underestimate of the real metallicity \citep{peimbert_temperature_1967, mendez_delgado_t_inhomogeneities_2023}. On the other hand, there also are indications, based on the comparison with fine-structure infrared lines such as $\mathrm{[O III] 52,88 \mu m}$, that temperature fluctuations might not be the main responsible for the ADF \citep{Tsamis_ADF_2003, Chen_Mrk71__2023}. In fact, the presence of small-scale, chemical inhomogeneities due to a clumpy, not well mixed gas distribution might possibly explain the different results arising from RLs (dominated by the metal-rich clumps) and the CELs (dominated by more diffuse medium) measurements \citep{Stasinska_abundance_bias_2007}.
Furthermore, spectroscopy performed on individual, supergiant stars in nearby spiral galaxies suggest that $\mathrm{T_e}$-based abundances are in better agreement with stellar metallicities (once taking into account the depletion onto dust) than those measured from recombination lines, especially in metal-poor environments \citep{bresolin_young_2016}.

\subsubsection{Statistical (strong-lines) methods}
\label{sssec:strong_line_methods}

Auroral lines involved in the classical $\mathrm{T_e}$-method are generally faint and their emissivity is strongly temperature-dependent, making them extremely challenging to detect, especially in high-metallicity systems.
This practically limits the applicability of `direct' metallicity measurements to a relatively small sample of galaxies, which either low-metallicity and/or high signal-to-noise spectra brings such faint emission line fluxes within the reach of current facilities.
Although the number of spectra with auroral lines detections have increased significantly (also at high redshift) over the past years thanks to the improvement in the astronomical instrumentation, they  still represent a tiny fraction of the global galaxy population.
Therefore, alternative methods need to be developed to study the metallicity properties of large galaxy samples like those probed by extensive spectroscopic surveys.

These techniques generally employ the (either direct or secondary) dependence of some line ratios between bright emission lines and metallicity (or, more specifically, oxygen abundance), and are referred to as the `strong-line methods'. Such a dependence can derive from the correlation between metallicity and electron temperature of the gas, or be driven by more indirect correlations between the metallicity and ionization parameter or through the relative abundances of different chemical species (like the nitrogen-to-oxygen ratio, e.g. \citealt{perez-montero_impact_2009}).
Such `strong-line' diagnostics are generally calibrated against `direct' oxygen abundance measurements in sample of individual galaxies (or exploiting spectral stacking, e.g. \citealt{curti_new_2017}) for which the $\mathrm{T_e}$-method can be adopted. When applied to different galaxy spectra they must be therefore interpreted in a `statistical' sense, and the inferred metallicity should be considered reliable only if the physical properties of the target match those of the calibration sample. 

The majority of the most popular strong-line diagnostics involves rest-frame optical line ratios, leveraging the presence of many bright transitions in this wavelength range and its accessibility.
Among multiple diagnostics proposed and calibrated in the past, some relies uniquely on a combination of oxygen and hydrogen lines (such as $\mathrm{R_{23} = ([O III]\lambda\lambda4959,5007+[O II]\lambda\lambda3727,29)/H\beta}$ and $\mathrm{R_{3} = [O III]\lambda5007/H\beta}$) and are thus primarily sensitive to the oxygen abundance, though showing a double-branched behavior due to the different relative importance of optical CELs in regulating the ISM cooling in different metallicity regimes\footnote{At high metallicities, the most efficient cooling channel is represented by the infrared $\mathrm{[O III]~88 \mu m}$ emission, which lowers the temperature below the excitation threshold of optical CELs. Thus, since Te decreases with increasing O/H, $\mathrm{[O III]\lambda5007/H\beta}$ decreases with oxygen abundance in this regime. At lower metallicities instead, Te is higher, powering the collisional excitation of the $\mathrm{[O III]\lambda5007}$ line, which becomes the dominant coolant: this represents an intermediate regime where diagnostics such as $\mathrm{R_{3}}$ and $\mathrm{R_{23}}$ are almost independent of O/H. Finally, at very low metallicity cooling is dominated by recombination, and collisional excitation of $\mathrm{[O III]\lambda5007}$ scales proportionally with O/H.} \citep{pagel_composition_1979, mcgaugh_h_1991}.
Several authors hence proposed to use combinations of multiple diagnostics to break the degeneracy in the metallicity solution, leveraging the monotonic (though often `secondary') dependence on oxygen abundance of line ratios such as $\mathrm{O_{32} = ([O III]\lambda5007/[O II]\lambda\lambda3727,29)}$ (sensitive primarily to the ionization parameter, \citealt{kobulnicky_metallicities_2004}), $\mathrm{N_{2}=[N II]\lambda6584/H\alpha}$ and $\mathrm{O_3N_2 = R_3 - N_2}$ (which introduce a direct dependence on the nitrogen abundance, hence their interpretation as oxygen abundance tracers is sensitive to the distribution in N/O of the calibration sample, \citealt{pettini_oiiinii_2004}). 
Additional diagnostics involving emission lines of different elements leverage the proximity in wavelength to reduce the impact of uncertainties on the reddening correction, and include e.g. $\mathrm{Ne_3O_2 = [Ne III]\lambda3869/[O II]\lambda\lambda3727,29}$ \citep{nagao_gas_2006}, $\mathrm{N_2S_2H\alpha = [N II]\lambda6584/[S II]\lambda\lambda6717,31 + 0.22~[N II]\lambda6584/H\alpha}$ \citep{dopita_chemical_2016}, $\mathrm{S_{2}= [S II]\lambda\lambda6717,31/H\alpha}$ and $\mathrm{R_3S_2 = R_3 + S_2}$ \citep{curti_massmetallicity_2020}.
Each of these diagnostics carries both advantages and disadvantages, related to the relative strength of the lines involved, their wavelength separation (and hence reddening sensitivity), and dependence on other chemical elements or physical parameters rather than oxygen abundance: therefore, combining multiple diagnostics represents often a valuable approach to marginalize over these different effects.
For a more thorough discussion on strengths and weaknesses of the hereby discussed (and more) strong-line diagnostics, we refer to the review by \cite{maiolino_re_2019}.

\subsubsection{Photoionisation models}
\label{sssec:phot_models}

Beside strong-line diagnostics calibrated `empirically' over samples of observed galaxy spectra, extensive development in the area of photoionisation modeling over the past years culminated in a broad set of theoretical frameworks aimed at determining the physical properties of galaxies (including metallicity) via simultaneous fitting of multiple emission line features.
The classical approach is to use photoionization codes such as \textsc{CLOUDY} \citep{ferland_cloudy_2017} or \textsc{mappings} \citep{dopita_shock_1996}, possibly adopting different types of input ionizing radiation, to generate multiple grids of emission line ratios whose dependence upon gas properties such as ionization parameter\footnote{The ionization parameter is defined as the dimensionless ratio of the number of ionizing photon emitted per unit time $\mathrm{Q_{ion}}$ to the gas density, normalized by the speed of light, i.e. $\mathrm{U = Q_{ion}/(4\pi r^2n_e c)}$, where $r$ is generally calculated as the distance to internal surface of the illuminated cloud.} and gas-phase metallicity can be fully modeled and explored \citep[e.g.][]{kewley_using_2002, dopita_new_2013, gutkin_modelling_2016}.

More recently, this approach has been frequently incorporated into full SED fitting codes (already discussed in Section~\ref{sssec:full_sed}) to simultaneously constrain stellar and nebular properties within robust Bayesian frameworks.  
The effect of dust is also generally included, both in terms of attenuation (sometimes also distinguishing the contribution from the dense clouds where stars are born from the that of the diffuse ISM, e.g. \citealt{chevallard_beagle_2016}) and of differential depletion of chemical elements.
The flexibility of these codes allows to probe a large number of galaxy properties over wide dynamic ranges, beyond those typically accessible by the majority of empirical calibration samples (especially at high redshift), and to have a good handle on the parameters' uncertainties, degeneracies, and correlations, although being intrinsically tied to the assumptions of the underlying model grids.

Despite the very high potential in fact, photoionisation models suffer from a number of limitations associated for instance with over-simplified geometrical configurations (with $\mathrm{HII}$ regions often modeled as plane-parallel slabs of gas), relative chemical abundances assumed to scale proportionally to the solar pattern or by means of fixed relationships with no-scatter, uncertainties on the input stellar libraries (though extensive work has been done recently to include for instance the effect of binaries and stellar rotation, e.g. \citealt{Stanway_Eldrige_BPASS_2018}), and constant dust-to-metal ratios with both metallicity and redshift.
This often translates into the inability for a given model to match simultaneously all the observed line ratios for an individual galaxy with the same, fixed set of parameters, although the statistical distribution of galaxy samples within specific diagnostic diagrams is generally well reproduced.

Furthermore, the metallicity inferred from photoionization models is generally overestimated compared to that derived by means of more `direct' methods such as the $\mathrm{T_e}$-method, with discrepancies that can reach up to $\sim0.6$~dex, especially in the high metallicity regime \citep{kewley_metallicity_2008}. 
Among the possible sources of discrepancies, there are a different treatment of dust depletion (which models generally account for, whereas `direct' measurements only probe metals in the gas-phase) and temperature fluctuations in real nebulae (see Section~\ref{sssec:te_method}), possibly producing an underestimate of the `true' abundances with the $\mathrm{T_e}$-method on the one hand, as well as some of the simplistic models' assumptions discussed above (for instance related to the scaling of nitrogen-to-oxygen abundance, \citealt{nicholls_abundance_2017}) on the other.  
Regardless of which can be deemed as the `true' absolute values (although $\mathrm{T_e}$-measurements are generally considered more physically motivated, as inferred from real observations of nebular spectra and descending from basic concepts of atomic physics), it is warranted to stress the importance of sticking to the same, self-consistent abundance scale when performing comparative studies in the literature, as hybrid combinations of model- and $\mathrm{T_e}$-based calibrations can introduce subtle systematics effects as a function of metallicity which might hamper the correct interpretation of the observations.  

\subsection{Absorption line techniques}
\label{ssec:absorption_lines}

If a gas cloud is observed against a bright background source of radiation (e.g. a distant quasar), the absorption produced by a given transitions provides a direct measurement of the column density of the associated ionic species.
Combined with a measurement of the hydrogen column density through one of the Lyman absorption lines, and modulo properly accounting for the different ionization states of the gas, it is possible to obtain very accurate measurements of the metallicity as well as of relative chemical abundances of different elements.
These techniques offer several advantages, including being sensitivity to extremely low column densities (down to $\mathrm{N\simeq 10^{12}~cm^{-2}}$), and the possibility to probe a wide range of densities (unlike emission-lines which intensity scales as $\mathrm{\approx N^2}$), redshifts, and target luminosities.
On the other hand, absorption studies provides only projected measures of the gas surface density, and are usually limited to one sightline per galaxy given the rarity of background quasars. 

Absorption line techniques can probe systems across a large range of column density regimes.
Up to column densities of $\mathrm{log(N/cm^2)}\approx16$, absorption profiles can be studied by directly measuring the equivalent width of the line, whereas at higher densities saturation effects starts to become relevant, often providing only lower-limits on the inferred column density of the ionic species and requiring to fit high-order Lyman series lines to derive the hydrogen column density.
Systems with $\mathrm{log(N_{HI}/cm^{-2})>16.2}$ are referred to as `Lyman limit systems' (LLS), as they cause a clear drop-off in the measured flux at the Lyman limit of $912\AA$, and typically probe the dense part of the CGM of galaxies (see Section~\ref{ssec:cgm}).
Above $\mathrm{log(N_{HI}/cm^{-2})\gtrsim 20.3}$, the highest Lyman series lines are saturated, and column densities must come from fitting of the Ly$\alpha$ profile, which on the contrary starts showing clear signatures of damped wings: these are so-called `damped Lyman-$\alpha$' (DLA) systems, and represent the class of systems most widely exploited to trace metal enrichment through absorption features at cosmological distances \citep[e.g.][]{Pettini_abundances_book_2004, Poudel_DLA_2018}.
It is worth noting that, even in those cases in which the optical counterpart of the absorber is identified, it is not straightforward to associate the DLA system with a specific region of a galaxy, e.g. whether it is tracing the ISM in the outer disks or dense gas clumps within the galaxy's circumgalactic medium (CGM). 
Nonetheless, the process of associating absorbing systems with emitting galaxies have made huge progress in the last decade, especially thanks to last generation IFU instruments such as \textsc{VLT/MUSE} \citep[e.g.][]{lofthouse_MUSE_MAGG_2020} and \textsc{KECK/KCWI} \citep[e.g.][]{martin_kcwi_inflows_2019}.

Since these observations generally probe atomic hydrogen and a fraction of the metal ionic states along any given line-of-sight, sometimes significant ionization corrections are required to move from a measured ratio of ionic to $\mathrm{HI}$ column density $\mathrm{N(X^i)/N(H^0)}$ to an abundance ratio $\mathrm{[X/H]}$. 
Ionization correction factors are generally estimated from photionisation codes like \textsc{Cloudy} \citep{ferland_cloudy_2017}, assuming simple geometrical configurations with an ionizing radiation field incident on a plane-parallel slab of constant density gas.
At the large column densities probed by DLA systems however, the optical depth is large enough for the interior of the cloud to be self-shielded from the external radiation field, hence the gas is predominantly in the neutral phase and chemical abundance measurements require little ionization corrections. 

Finally, a robust measurement of the metal content with absorption line techniques also requires to account for depletion of metals onto dust grains.
This process is differential, as some elements are more likely to be incorporated into dust grains based on their chemical properties: for instance, over $50\%$ of metals in the Milky Way’s ISM are in the solid phase, but elements like Fe are extremely depleted with up to $90-99\%$ of the total abundance incorporated into dust. Neutral absorbers show differential depletion patterns similar to, but less severe than, those in the Milky Way. 
Evidence of metal depletion extends to high-redshift systems, and indicators of dust presence, such as the $2175\AA$ absorption bump \citep{witstok_UVbump_2023} and strong infrared absorption features \citep{aller_silicate_dust_absorber_2014}, further confirm the presence of solid-phase material in these environments.
While early studies focused on lightly depleted elements, such as zinc (Zn) and silicon (Si), to estimate total metal abundances \citep[e.g.][]{Pettini_zinc_DLA_1997}, more recent works have used empirically-calibrated depletion patterns to correct gas-phase measurements, deriving more accurate total metal abundances \citep{jenkins_depletion_2009, decia_dust_2018}.

\section{Chemical Evolution Models}
\label{sec:CHEMs}

The chemical enrichment in galaxies is regulated by the complex interplay of multiple physical processes occurring on different spatial and temporal scales.
These include gas accretion from the cosmic web, stellar evolution and nucleosynthesis, stellar winds and SNe explosions, gas outflows (possibly removing gas from the galaxy), gas recycling from the surrounding CGM, dust creation and destruction, AGN activity, dynamical interactions and mergers.
Over the past decades, detailed analytical and numerical frameworks aimed at describing the impact (and relative role) of all these processes in driving galaxy evolution and, more specific to this context, at predicting the observational signatures in their chemical enrichment patterns, have been extensively developed. 
Despite a variety of different approaches, assumptions, and levels of complexity, most of these models lie on the foundation of simple equations describing galactic chemical evolution (GCE), and involve a number of common relevant parameters, as described in the following.

\subsection{Analytical models}
\label{ssec:analytical_models}

Analytical models of galactic chemical evolution generally attempt to describe the interplay between four main processes, described by means of as many parameters: gas accretion, the efficiency of gas conversion into stars (star formation efficiency), the amount of metals produced by stars and returned to the ISM (stellar yield), and the amount of gas lost from the system due to stellar winds and outflows.
Of particular relevance is the definition of stellar yields, which are derived from models of stellar evolution.
The yield is defined as the fraction in mass of a given element produced per unit mass of stars formed and returned to the ISM; yields therefore depend on the mass of the star, its chemical composition, and other properties such as stellar rotation. For instance, a supernova originating from a $\mathrm{35~M_\odot}$ star produces $\approx \mathrm{5~M_\odot}$ of oxygen and $\approx \mathrm{1~M_\odot}$ of carbon \citep{Chieffi_yields_2004}.
When modelling GCE, yields are computed for full simple stellar populations (i.e. for stars formed at the same time) by integrating the yields of individual stars over the initial mass function (IMF). These yields depend on the ISM enrichment at the time of the formation of the stellar population and, possibly, also on time, to reflect any cosmic evolution in the IMF. Throughout the simple treatment of GCE within this Chapter, we will work under the assumption of a constant IMF.

In classical models, the equations describing a galactic system can be simplified under the assumption that stars can be divided into two classes: those that live forever (with $\mathrm{M_\star}$ below some threshold, $\mathrm{M_\star < M_0}$) and those that die out as soon as they are born ($\mathrm{M_\star > M_0}$): this is known as `instantaneous recycling approximation' (IRA).
Such approximation is justified by the fact that most of the metals are produced by Type II SNe of massive stars progenitors ($\mathrm{\gtrsim 8 M_\odot}$) with short lifetimes of $\approx$~a few Myrs.
However, these models fall short in capturing the full chemical evolution of elements like iron (Fe), whose enrichment occur on much longer timescales ($\approx$Gyrs) due to significant contribution from type Ia SNe.

Under the IRA conditions (and assuming the metallicity $Z$ is always $\ll 1$, hence the mass in metals can be ignored compared to the total gas mass), the basic equations describing galactic chemical evolution can be expressed as : 

\begin{align}
\label{eq:IRA}
    \mathrm{
    \frac{dM_s}{dt} = (1 - R) \psi \ ; 
    \frac{dM_g}{dt} = -(1 - R)\psi + \phi - \omega \ ;
    \frac{d(ZM_g)}{dt} = -Z(1-R)\psi + y(1 - R) \psi + Z_{\phi} \phi - Z \omega \ ;
    }
\end{align}
where $\mathrm{M_s}$ is the mass in stars, $\mathrm{M_g}$ is the mass in gas, $\mathrm{\psi}$ is star-formation rate (i.e. the amount of stellar mass formed per unit time), $\mathrm{\phi}$ is the gas infall rate, $\mathrm{\omega}$ is the outflow rate (with both $\mathrm{\phi}$ and $\mathrm{\omega}$ which are functions of time), $\mathrm{R}$ is the `return fraction' (i.e. the fraction in mass returned to the ISM per stars formed, which depends on the IMF), $\mathrm{y}$ is the yield, $\mathrm{Z}$ is the metallicity of the ISM, $\mathrm{Z_{\phi}}$ is the metallicity of the infalling gas, while the metallicity of the outflowing gas is the same as that of the ISM.
Generally, the last equation can be written explicitly in terms of the metallicity $\mathrm{Z}$ (instead of the total mass of metals in the gas-phase, $\mathrm{ZM_g}$), 
\begin{align}
\label{eq:closed_box1}
        \mathrm{
        M_g\frac{dZ}{dt} = y(1 - R) \psi + (Z_{\phi} - Z)\phi \ .
        }
\end{align} 

\paragraph{\textbf{The closed box model}}
The simplest realisations of GCE models assume that all the gas is already present initially and is pristine (no gas inflows, $\mathrm{M_g (t=0)=M_0}$, $\mathrm{Z (t=0)=0}$), that metals are produced, returned, and mixed into the ISM, and that this enriched gas forms the next generations of stars without any loss of gas (no outflows): these are generally referred to as `closed box models', and explore the evolution of only three quantities, i.e. the gas mass $\mathrm{M_g}$, the stellar mass $\mathrm{M_s}$, and the metallicity Z.
Since the total mass in the systems remains constant with time, i.e. $\mathrm{M = M_s + M_g =M_0}$, integrating the equations~\ref{eq:IRA} and \ref{eq:closed_box1} leads to
\begin{align}
\label{eq:closed_box2}
    Z = y \ln \mu_g^{-1} \, 
\end{align}
where $\mathrm{\mu_g \equiv M_g/(M_g + M_s) = M_g/M_0}$ is the `gas fraction' of the system. Therefore, in this model the metallicity increases monotonically as the gas mass decreases due to star formation.
To fully solve the closed box model we need to specify how efficiently gas is converted into stars, which we parametrize as the `star-formation efficiency' $\mathrm{\tau = M_g/\dot{M}_s}$, 
here assumed constant over time\footnote{Note that a larger value of $\tau$
corresponds to a longer time required to consume the available gas, hence to a lower efficiency of star formation}.
Because at any given time the variation in the gas mass corresponds to an opposite variation in the stellar mass (i.e. $\mathrm{\dot{M_g}=-\dot{M}_s}$), we obtain the following expressions describing the temporal evolution of gas mass and star-formation rate
\begin{align}
\label{eq:closed_box3}
    \mathrm{
    M_g(t) = M_g (0)\ e^{-t/\tau} \ ; \  \dot{M_s}(t) = \frac{M_g (0)}{\tau}\ e^{-t/\tau} \ .
    }
\end{align}
The metallicity evolution is then given by 
\begin{align}
\label{eq:closed_box_4}
\mathrm{
Z(t) = -y \cdot \mathrm{ln} \Bigg [ \frac{M_g (t)}{M_g (0)} \Bigg ] = y \ \frac{t}{\tau} \ .     
}
\end{align}
From equation~\ref{eq:closed_box_4} we see that the metallicity reaches its peak $Z=y$ at $t=\tau$. 
Applying this scenario to the Milky Way, assuming the present-day measured gas fraction of the Solar neighborhood $\mu_g$=0.1 and a solar metallicity $Z_\odot=0.014$ \citep{asplund_solar_2009}, we derive a yield $y\sim0.006$ and a timescale for peak metallicity of $\mathrm{t=\tau \ Z_\odot/y \approx 2.33 \tau}$.

Comparing the closed box model predictions with observations however immediately highlights several critical aspects of its oversimplified approach. 
For instance, the stellar yield value inferred as above is significantly lower than the typical yields computed from stellar evolution at solar metallicity ($y\sim0.035$). Furthermore, the metallicity distribution predicted by the closed box model includes a much larger number of low-metallicity stars than observed in the solar neighborhood: in fact, the model predicts a mass fraction of stars of metallicity below $25\%$ solar of $\mathrm{M_s[Z<0.25Z_\odot] \ \sim 0.45}$, whereas the observed value is $\sim 0.02$.
The origin of such disagreement (also known as the `G-dwarf problem') resides in the fact that early star formation and enrichment produces an amount of metals which is diluted by the large gas reservoir required to sustain star-formation throughout the life of the system (indeed, we recall that in closed box models all the gas is immediately available to form stars). 
Therefore, the metallicity initially grows slowly, and many stars therefore form from low-metallicity gas for a long time. As the gas reservoir starts to get exhausted with time, star formation produces the same yield of heavy elements, which however enrich an ever smaller amount of gas such that significantly fewer stars form in relatively high metallicity ranges. 
A possible solution to this problem is to model galaxies as `open systems', including the contribution from gas inflows and outflows in our prescriptions.

\paragraph{\textbf{The impact of gas flows}}
It is relatively straightforward to understand that galaxies are far from being `closed boxes'. On the one hand in fact, they accrete gas from their cosmic environments, while on the other gas can be removed by winds driven by either massive stars, supernova explosions, and active galactic nuclei.

Typically, the amount of gas lost due to outflows per unit time (the $\omega$ term in equation ~\ref{eq:IRA}) is parameterized in terms of the `mass loading factor' $\eta$, expressed as a fraction of the star-formation rate $\mathrm{\psi \equiv \dot{M_s}}$
\begin{align}
\label{eq:leaky_box1}
    \mathrm{
   \omega \equiv \dot{M}_{\text{outflow}} = \eta \ \dot{M}_s \ ,
   }
\end{align}
which implicitly assumes that outflows are primarily driven by the winds or supernova explosions of massive stars.
Incorporating outflows in the equations of the closed box model, we obtain
\begin{align}
\label{eq:leaky_box2}
    \mathrm{
    \dot{M}_g = -\dot{M}_s - \dot{M}_{\text{outflow}} = -(1 + \eta) \ \dot{M}_s \ 
    }
\end{align}
for the evolution of the gas mass, whereas the amount of mass in metals $M_Z$ evolves as
\begin{align}
\label{eq:leaky_box3}
\mathrm{
    \dot{M}_Z = (y - Z) \ \dot{M}_s - Z \ \dot{M}_{\text{outflow}} = (y - Z - \eta Z) \ \dot{M}_s \ .
    }
\end{align}
Combining equations~\ref{eq:leaky_box2} and \ref{eq:leaky_box3}, and expressing everything in terms of metallicity Z, we obtain
\begin{align}
    \mathrm{
    \frac{dZ}{dt} = - \frac{y}{1+\eta} \  \frac{\dot{M}_g}{M_g} \ ,
    }
\end{align}
hence
\begin{align}
\mathrm{
 Z(t) = - \frac{y}{1+\eta} \ \mathrm{ln} \Bigg [ \frac{M_g (t)}{M_0} \Bigg ] \equiv y_\text{eff} \  \mathrm{ln} \  \mu_g ^{-1} \ ,
 }
\end{align}
where we have defined the `effective yield' $\mathrm{y_\text{eff} = y/(1+\eta)}$ in order to recover the same functional form for the evolution of the metallicity $Z$ as in the closed box model (cfr. equation~\ref{eq:closed_box2}): here, the effective yield is lower by a factor $1+\eta$ than the metal yield of the stellar population to account for the fraction of enriched gas expelled from the galaxy.
This class of models takes the name of `leaky box models' and allow, for example, to reconcile the predicted and measured values of the stellar yields (by taking $\eta \approx 5$), as well as to bring the expected number of low-metallicity stars in better agreement with observations.
The inclusion of gas outflows in more detailed realisations of GCE models allows to successfully reproduce some of the most relevant scaling relations observed in galaxy populations, such as the mass-metallicity relation \citep[][see Section~\ref{sssec:MZR}]{tremonti_origin_2004}, which is indeed predicted to be driven by a decreasing efficiency of outflows in higher mass galaxies (characterised by increasingly higher escape velocities due to their deeper gravitational potentials) leading to increasing metallicity as a function of stellar mass.

Alternatively (or in addition), we can incorporate a term which reflects the accretion of gas from the cosmic web providing the supply for star-formation.  
Indeed, many cosmological frameworks predict that gas accreted in the form of cold streams is the main driver of galaxy assembly \citep{dekel_cold_2009}.
Including gas accretion into the equations of closed box models give rise to the so-called `accreting box models'. In their simplest form, we can assume that the accretion of gas perfectly balance the gas consumed by star-formation rate ($\phi=(1-R) \ \psi$), such that the total gas mass remains constant with time ($M_g = M_0$).
Substituting this into equations~\ref{eq:IRA}, and under the additional assumption of pure pristine gas accretion (i.e. the metallicity of the infalling gas is equal to zero), we obtain
\begin{align}
    \mathrm{
    \dot{Z} = (y - Z) \ \frac{\dot{M_s}}{M_g} \ (\text{since} \  \dot{M}_g=0) \ 
    }
\end{align}
and, by means of a change of variable from time to the total mass of the system $M=M_g + M_s$, and considering that $\mathrm{\dot{M} = \dot{M}_s}$, we have 
\begin{align}
    \mathrm{
    \frac{dZ}{dM} = \frac{\dot{Z}}{\dot{M}_s} = \frac{(y-Z)}{M_g} \ .
    }
\end{align}
In the scenario of zero initial metallicity ($Z(0)=0$) and $M(0)=M_g$, a possible solution is in the form
\begin{align}
    \mathrm{
    Z = y \  \Bigg [ 1 - \mathrm{exp} \  \Bigg ( 1 - \frac{M}{M_g} \Bigg ) \Bigg ] \ .
    }
\end{align}
In such a model, the total mass quickly becomes larger than the gas mass right after the beginning of star-formation (which is fueled by external accretion), and for $M \gg M_g$ the metallicity of the system reaches $Z\approx y$, representing the equilibrium value between enrichment and pristine accretion which is directly set by the stellar yield.
The accretion box models provides better agreement between both the inferred ($y\approx0.014$\footnote{ assuming the Milky Way's present-day gas fraction and solar metallicity}) and observed ($y\approx0.035$) yields as expected from stellar evolution, as well as for the fraction of low-metallicity stars in the solar neighborhood.

We have seen therefore how including the contribution of gas flows (in the form of pristine cosmological gas accretion and/or gas removal from outflows) into a closed box system helps to reconcile models predictions with observations.
In general, advanced analytical chemical evolution models account for both processes in their frameworks.
A widely adopted class of models assume that gas inflow is compensated by star formation and outflows, yielding a quasi-steady state in which the gas content only slowly varies with time: these are hence known as `equilibrium' or `gas-regulator' models \citep[e.g.][]{peeples_mzr_model_2011, lilly_gas_2013}.
These models have proven successful in reproducing multiple average properties of galaxy populations, in particular the metallicity scaling relations, which in turn provide some of the most relevant observational constraints to the models' parameters and, in turn, to the mechanisms regulating star-formation and mass assembly in galaxies that such parameters describe.

\subsection{Numerical models and simulations}
\label{ssec:simulations}

These models simulate the hierarchical growth of cosmic structures via cosmological accretion and merging, starting by N-body simulations tracking the effects of gravity on both dark and baryonic matter. 
Key elements in these models are the spatial (and mass) resolution and the total size of the simulation, the latter related to the total number of particles and the volume of the Universe being probed. 
Due to highly demanding computational resources of these simulations, it is almost never possible to simultaneously study larger volumes at high-resolution.
Therefore, although in many cases the share the same underlying physical principles, different simulations are run probing a range of different resolutions and sizes depending on their primary scientific objectives.
Among notable examples of large scale, cosmological simulations of galaxy formation and evolution implementing also chemical enrichment processes, we mention ILLUSTRIS \citep{Vogelsberger_ILLUSTRIS_2014}, ILLUSTRIS-TNG \citep{nelson_TNG_2019}, EAGLE \citep{Schaye_EAGLE_2015}, HORIZON-AGN \citep{Dubois_HORIZON_AGN_2014}, FIRE \citep{Hopkins_FIRE_2014}, and MUFASA \citep{dave_mufasa_2017}, whereas hybrid, semi-analytical approach is followed by the L-GALAXIES simulations \citep{Henriques_LGALAXIES_2020}.
Extremely detailed, high-resolution simulations of individual galaxies are instead provided e.g. by SERRA \citep{pallottini_zooming_2017} and RAMSES \citep{Katz_RAMSES_2022}.
For a more comprehensive overview we refer to \cite{somerville_dave_2015}.

\section{The observational landscape}
\label{sec:observations}

Few research areas in astronomy have experienced the same progress over the past two decades as the study of the chemical evolution of galaxies.
Major advancements in observational techniques and the advent of both ground- and space-based cutting-edge facilities with high sensitivity, high multiplexing, and wide wavelength coverage have in fact enabled to probe the chemical evolution in large galaxy samples, map their metal content, and directly probe the enrichment process of galaxies and of the intergalactic/circumgalactic medium up to the earliest cosmic epochs, setting tight constraints on models of galaxy formation.
In this Section we provide a brief overview of the current 
observational landscape regarding the study of the metallicity scaling relations, the relative abundances of chemical elements, and their spatial distribution in and out of galaxies, trying to connect these results over a wide range in cosmic history.

\begin{figure}
    \centering
    \includegraphics[width=0.95\linewidth]{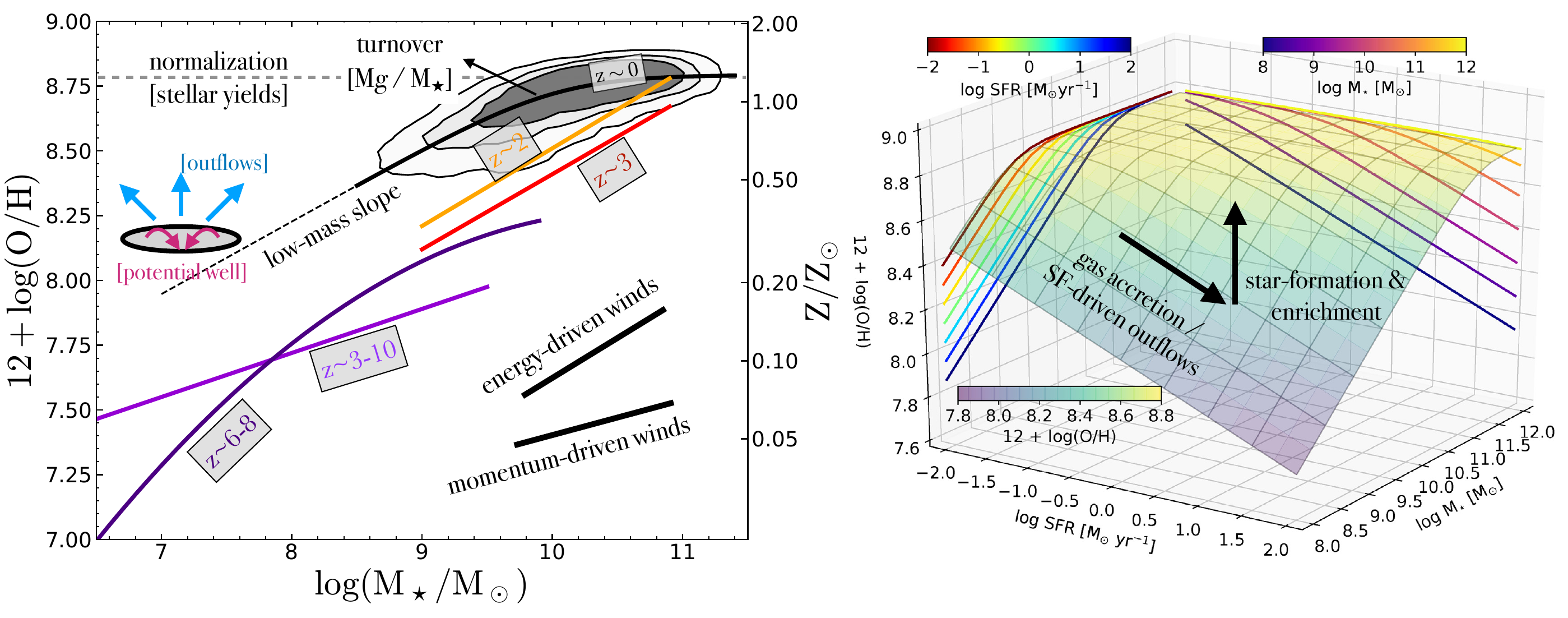}
    \caption{Gas-phase metallicity scaling relations. Left Panel: Mass-Metallicity Relation (MZR) as observed in the local Universe (grey contours, $\mathrm{ z \sim 0.08}$, \citealt{curti_massmetallicity_2020}), and for different samples at high-redshift ($\mathrm{ z \sim 2-3}$ from \citealt{sanders_mosdef_mzr_2021}; $\mathrm{ z \sim 3-10}$ from \citealt{curti_jades_mzr_2024}; $\mathrm{ z \sim 6-8}$ from \citealt{Chemerynska_MZR_uncover_2024}).
    Some of the physical processes responsible for shaping the relationship are schematized, and the slopes predicted by  assuming either `energy-driven' or `momentum-driven' winds are also overlaid (with arbitrary normalization).
    Right Panel: Graphical representation of a possible parametrisation of the FMR surface in the 3D space given by $M_\star$, SFR, and O/H. The colour-coding of the surface reflects the predicted metallicity, and the contours of the projections of the FMR onto the $M_\star$-O/H and SFR–O/H planes are also drawn. Adapted from \citealt{curti_massmetallicity_2020}.
    }
    \label{fig:scaling_relations}
\end{figure}

\subsection{Metallicity scaling relations}
\label{ssec:scaling_relations}

The incessant interplay between gas flows, star-formation, and metal enrichment that regulates the baryon cycle leaves clear signatures on both stellar and ISM metallicity. Observationally, this is reflected into the existence of scaling relations between metallicity and other global galaxy properties, whose investigation can help shedding light onto the relative role that these processes have in driving galaxy evolution. 

\subsubsection{The Mass-Metallicity Relation}
\label{sssec:MZR}

The most evident and thoroughly studied scaling relation relates the total stellar mass with the metal content in galaxies, and is known as the `Mass-Metallicity Relation' (MZR), and it is observed for both stellar and gas-phase metallicity.
Observationally, the MZR in the gas-phase is a rather tight correlation ($\sim 0.1$~dex scatter) in which the more massive systems appear more metal enriched. 
Initially suggested in the form of a correlation between metallicity and luminosity in irregular and blue compact galaxies \citep{lequeux_chemical_1979}, it was then fully characterised with the advent of large spectroscopic surveys like the Sloan Digital Sky Survey \citep[SDSS][]{york_sloan_2000}, which allowed to study the MZR over three orders of magnitude in stellar mass \citep{tremonti_origin_2004, andrews_mass-metallicity_2013, curti_massmetallicity_2020} and even down to the regime of dwarf galaxies \citep{lee_extending_2006, yates_present-day_2019}, despite such low-mass regime being more prone to sample selection biases \citep{Scholte_MZR_DESI_2024}.

The MZR presents a relatively steeper slope at low and intermediate masses, while evidence for a turnover and flattening in the high-mass regime is observed.
The main features of the MZR, i.e. the slope, turnover mass, and normalisation, can be linked to different physical processes, and their characterisation provides some of the most powerful constraints on galactic chemical evolution models (left-hand panel of Fig.~\ref{fig:scaling_relations}).
The MZR slope reflects feedback mechanisms in galaxies as shaped by star-formation-driven outflows and, in particular, how the mass-outflow-rate scales with the stellar mass \citep{peeples_mzr_model_2011, sanders_mosdef_mzr_2021}. 
The gas expelled is generally more enriched than the average ISM, and outflows are expected to be much more effective in low-mass galaxies with shallower gravitational potential wells \citep{tremonti_origin_2004}.
The normalisation of the MZR (whose exact determination is however largely dependent on the assumed method for metallicity determination, see Section~\ref{sec:met_measurements}) reflects instead the asymptotic  metallicity value reached in equilibrium conditions, i.e. when the amount of metals produced by massive stars equals the mass of metals locked up in low-mass stars, and is therefore directly related to the nucleosynthetic stellar yields.
Finally, in several analytical models the `turnover' mass (i.e. the stellar mass value at which the MZR begins to flatten-out) is related to the gas fraction in galaxies \citep{zahid_universal_2014}.

Alternative and complementary mechanisms invoked to explain the shape of the gas-phase MZR include on-going accretion of pristine gas (diluting the metal content of the ISM, \citealt{lagos_fundamental_2016}) and re-cycling of previously enriched gas from the CGM, the latter more relevant in high-mass systems \citep{peroux_2020}.
Moreover, the shape of the high-mass end of the IMF could vary with galaxy mass, introducing a systematic change in the average stellar yields and in the rate of metal enrichment \citep{vincenzo_modern_2016}.
Another possible interpretation of the origin of MZR invokes the so-called `downsizing' scenario of galaxy evolution, in which massive galaxies form earlier and evolve faster (being hence more `chemically mature') compared to low-mass systems: in this sense, the MZR represents therefore an evolutionary sequence \citep{somerville_dave_2015}. 
Finally, recent extensive analysis of large statistical samples of resolved MaNGA galaxies employed machine learning techniques to explore the underlying connection between gas-phase metallicity, stellar mass, and proxies of the galaxy gravitational potential, leading however to contradicting results regarding which physical parameter has to be considered the main driver of the MZR \citep{baker_mzr_2023, sanchez-menguiano_MZR_grav_pot_2024}.


A relationship between stellar mass and stellar metallicity in galaxies is also observed \citep[e.g.][]{gallazzi_ages_2005, zahid_stellar_2017}, although its shape and normalisation depends on whether the derived stellar metallicities are luminosity-weighted (probing mainly young stellar populations and hence being more similar to the gas-phase MZR) or mass-weighted (reflecting the integrated SFH of galaxies and mainly probing metals locked into older stellar populations). 
Comparing directly gas-phase and stellar MZR provides useful insights into the role of galactic outflows and possible variations in the IMF \citep{lian_mass-metallicity_2018}.
Furthermore, any offset between the stellar MZR of star-forming and passive systems can be related to the physical processes responsible for quenching star-formation in galaxies.
Such a difference is for instance clearly observed in local SDSS galaxies, with passive systems being significantly more enriched than star-forming ones of the same stellar mass \citep{peng_strangulation_2015, Trussler_starvation_2020}.
A possible scenario to explain this discrepancy invokes the so-called `strangulation' mechanism of galaxy quenching, in which gas accretion is halted in passive galaxies by either dynamical processes or feedback-heating of the galaxy halo while the metallicity keeps increasing as the available gas reservoir is totally consumed. 

\subsubsection{The relationship between mass, metallicity, star-formation rate, and gas content}
\label{sssec:FMR}

Despite its relative tightness, the scatter in the gas-phase MZR is observed to correlate with different galaxy properties. Observationally, the most relevant dependence is with the current level of star formation, in that galaxies with higher star-formation rates are observed to be more metal depleted, on average, than low star-forming galaxies of the same stellar mass.
Such secondary dependence of the MZR on the star-formation rate was initially reported by \cite{ellison_clues_2008}, and lately parametrised by \cite{mannucci_fundamental_2010} into a full, three-dimensional relationship between $M_\star$, SFR, and metallicity often referred to as the `Fundamental Metallicity Relation' (FMR, right-hand panel of Fig.~\ref{fig:scaling_relations}). 
By including the SFR into the framework, the scatter around the MZR is reduced down to a level ($\sim0.055$~dex) comparable with the typical measurement uncertainties \citep{mannucci_fundamental_2010, salim_critical_2014, curti_massmetallicity_2020}.
Several other parametrizations of the FMR have been proposed, as its exact shape depends significantly on both sample selection and methods adopted to measure the three relevant physical quantities \citep{yates_relation_2012, andrews_mass-metallicity_2013, telford_exploring_2016, hunt_coevolution_2016}.
Despite some initial contradicting claims \citep{sanchez_califa_2016, barrera-ballesteros_separate_2017}, this relationship has been observed to hold also at the local, spatially resolved level probed by integral-field spectroscopic surveys \citep{cresci_fundamental_2018}, revealing the effect of the interplay of processes occurring on both small scales (e.g. local mass assembly) and global scales (e.g. galaxy-wide outflows) onto the chemical enrichment of individual star-forming regions within galaxies \citep{barrera-ballesteros_galaxy_2016, baker_resolved_fmr_2023}.

The small amount of scatter around the FMR supports a scenario in which chemical enrichment is regulated by smooth, secular evolutionary processes which set an equilibrium between gas accretion, star-formation, and outflows.
Indeed, the existence of the FMR can be explained assuming that star-forming galaxies do most of their growing in a steady-state mode where  SFR, metallicity, and gas fraction constantly adjust on a relatively short timescale to reflect the influence of inflows. Cosmological accretion of external gas does not change the stellar mass, but it decreases the mean metallicity of the ISM while simultaneously triggering new star formation episodes. As time goes on, star formation consumes the gas and increases its metallicity, until new metal-poor gas arrives and the cycle starts over. If samples of galaxies with similar $M_\star$ are observed at different phases of this cycle (initially triggered by a gas accretion event) they will show a dispersion in metallicity which anti-correlates with the instantaneous SFR, producing the observational signature of the FMR. 
Theoretical frameworks such as `gas-regulator' models described in the previous section generally predict the existence of the FMR, which arises directly by the assumed balance between the inflow and outflow rates and galaxy internal processes \citep{dave_galaxy_2011, lilly_gas_2013, dayal_physics_2013}.
Similarly, an FMR is `observed' also in several cosmological simulations, as driven by the link between gas fraction and SFR \citep{de_rossi_galaxy_2017, dave_mufasa_2017, torrey_mzr_TNG_2019} and its exact shape also reflects the detailed prescriptions on the treatment of star-formation and AGN-driven feedback in galaxies \citep{ma_origin_2016}.

As suggested by even the more basic chemical evolution models, both SFR and metallicity are related to the gas reservoir of galaxies, in that higher gas fractions correspond simultaneously to higher SFR and lower metallicities at fixed $M_\star$,
and this principle is at the basis of the success of many analytical and numerical prescriptions in matching observations.
Therefore, it is not unexpected that such an anti-correlation between metallicity and gas fraction should be directly observed in galaxies and that, moreover, this might be even more `fundamental' (from a physical perspective) than that observed with the SFR.
Despite the large observational challenges associated with measuring the gas content in galaxies, both because of the systematics (and metallicity dependence) affecting its tracers or due to its multi-phase nature\footnote{hydrogen can be found in either atomic or molecular phase} \citep{bolatto_co--h2_2013, tacconi_phibss_2018}, several studies reported possible evidence of an anti-correlation between the gas-phase metallicity and either atomic \citep{bothwell_fundamental_2013, jimmy_gas_2015} or molecular \citep{bothwell_molecular_2016} hydrogen, the latter more directly related to the process of star-formation.

\subsubsection{Redshift evolution}
To study the evolution of the mass-metallicity relation with redshift is key to constrain the chemical yields of early stellar populations as well as to unveil the role played by gas flows in the regulation of the gas content in galaxies at different epochs and for different mass regimes.
Observationally, extensive spectroscopic campaigns have been conducted within the past decades, leveraging in particular the advent, from the early 2000s, of near-infrared multi-object spectrographs on 8-meters class telescopes that allowed to target rest-frame optical metallicity diagnostics in galaxies up to redshift of $z \sim3$. 

Despite the large variety of studied samples and adopted methodologies, a clear picture of the MZR evolution emerged, with high-redshift sources observed to be less enriched than local galaxies of the same stellar mass.
Such a decrease in galaxy metallicity with redshift at fixed mass can be ascribed to different mechanisms, including an enhanced gas fraction at high-z from large inflow rates, an increased efficiency in metals removal by star-formation-driven outflows, or a change in the stellar yields, possibly associated with an evolution in the shape of the stellar IMF.

The investigation of the redshift evolution of the MZR is prone to multiple
systematics effects, the most relevant being associated with biases in the excitation properties and star-formation rates of the selected samples, uncertainties on the reddening correction, and adoption of different sets of metallicity calibrations and diagnostics.
This led to discrepancies among literature studies in the determination of the `amplitude' of the evolution, both in terms of slope and normalisation.
While earlier works reported a large evolution, especially at low masses and $z\gtrsim2$ \citep[e.g. $\sim0.3-0.4$~dex at $\mathrm{M_\star=10^{10} ~M_\odot}$, ][]{maiolino_amaze_2008, mannucci_lsd_2009, onodera_ism_2016}, suggesting a `chemical downsizing' scenario\footnote{massive galaxies have formed faster, reaching the chemical enrichment level of their local counterparts at earlier epochs}, later studies based on more representative sample of $z\approx 1-3$ galaxies, and which adopted a careful and consistent approach in the metallicity determination via strong-line diagnostics, derived a relatively smooth variation in log(O/H) with redshift at fixed $\mathrm{M\star}$ and no clear evolution in the MZR slope in the `intermediate' mass regime ($\mathrm{9 \leq log(M\star/M_\odot) \leq 10.5}$) \citep{sanders_mosdef_mzr_2021, topping_mzr_2021}, with this trend now suggested to hold even in the dwarf galaxy regime \citep{Revalski_mzr_z1_2024}.
As the MZR slope reflects the efficacy of stellar winds in removing metals from the ISM, a redshift-invariant MZR slope up to $z\approx 3$ suggests that the physical mechanisms driving outflows in star-forming galaxies remain the same across more than $10$~Gyr of cosmic time \citep{sanders_mosdef_mzr_2021}.
At the same time, the evolution in the MZR normalisation towards lower O/H at fixed $M_\star$ with increasing redshift can be explained (beside the impact of metal-enriched gas outflows) by the `dilution' effect associated with the larger gas reservoirs of high-z galaxies, in agreement with the observed scaling of both gas content and star-formation efficiency with redshift \citep{tacconi_review_2020}, as well as with some of the most recent predictions from cosmological simulations based on hierarchical galaxy formation \citep{dave_mufasa_2017, torrey_mzr_TNG_2019}.  

When including the SFR into the picture however, hence moving to the framework of the FMR discussed in Section~\ref{sssec:FMR}, the situation changes.  
Remarkably, not only does accounting for the SFR dependence reduce the scatter of the $\mathrm{z\sim0}$ MZR, but high-redshift systems up to $\mathrm{z\sim2-3}$ have been observed to obey, on average, the same $\mathrm{z\sim0}$ \citep[e.g.][]{cresci_fundamental_2018, sanders_mosdef_mzr_2021}. This suggest that galaxy evolution proceeds as driven by the same interplay between gas flows and internal, secular processes over at least $\sim$~10~Gyr of cosmic time. 
Whereas the MZR is observed to evolve with redshift, the FMR concept captures well such scenario, explaining the MZR evolution as a projection effect driven by high-z galaxies preferentially probing different regions of the same, non-evolving FMR at different cosmic epochs \citep{mannucci_fundamental_2010}. Whether galaxy evolution in the very high-redshift Universe followed similar paths, and how this was reflected into the chemical enrichment properties of galaxies, have been unsolved questions for many years. 

\subsubsection{The advent of JWST and most recent developments}

The advent of the JWST marked the beginning of a new era by opening the long-awaited window onto the properties of high-z galaxies. Rest-frame optical lines previously inaccessible beyond $\mathrm{z\sim3}$ suddenly enabled to characterize the star-formation, metallicity, ionization, and dust properties of thousands of galaxies up to the earliest phases of structure formation.
Crucially, the improvement in sensitivity and wavelength coverage in the near-infrared provided by the NIRSpec instrument, coupled with the average decrease in metallicity (and enhancement in ionization) of early galaxy populations, allowed to perform much more robust metallicity measurements in high-z galaxies than ever before, either via direct detections of auroral lines in individual spectra \citep{curti_smacs_2023, laseter_auroral_jades_2023}, or exploiting the latter to deliver more reliable diagnostics for the high-z Universe \citep{sanders_calibrations_2023}.

At the time of writing this Chapter, and after more than two years of JWST operations, the landscape is in rapid evolution.
Nonetheless, the scenario depicted by early JWST studies reveals that the MZR is already in place by $z\sim6-8$ over a large range in stellar mass \citep{nakajima_mzr_ceers_2023, curti_jades_mzr_2024, Chemerynska_MZR_uncover_2024}, though with milder evolution in normalization from $z\sim2-3$ than several models would have predicted \citep[e.g.][]{ucci_astraeus_2023}, suggesting that star-formation and enrichment proceed fast in the early phases of galaxy assembly. Interestingly, indications of a possible flattening of the MZR slope in the low-$M_\star$ regime has been found \citep{curti_jades_mzr_2024}, which would point to a change in the main mechanisms driving galactic outflows and their scaling with stellar mass, favoring `momentum-driven' over `energy-driven' modes.

At the same time, and in tension with past observations out to $z\sim2-3$, a large fraction of high-z galaxies are observed to significantly diverge from the local FMR in that they appear less enriched than its local parametrization would predict for a given $M_\star$ and SFR \citep{Heintz_fmr_2023, nakajima_mzr_ceers_2023, curti_jades_mzr_2024}.
Furthermore, such a deviation appears not as a systematic offset but as a highly scattered distribution. This suggests that, although the global enrichment still reflects the assembled stellar mass (as seen in the MZR), the current SFR might not capture anymore second-order deviations in the metallicity, and hence the equilibrium between gas flows, star-formation, metal production and outflows that regulate galaxy growth across the latest $\sim10$~Gyr of cosmic time might be starting to break-down.
To understand whether such an offset can be reconciled by a change of parametrization (hence assuming a redshift-evolving FMR in which the relative strength of the metallicity-SFR dependence at fixed $M_\star$ changes with time, e.g. \citealt{Garcia_FMR_evolvI_2024}) or hint at more profound change in the main mode of galaxy assembly, will be a matter of future work.

Theoretically, deviations from the FMR at high-z could reflect different mechanisms in place. On one hand, they could be driven by cosmological processes in the form of a modulation of gas accretion within dark matter halos, possibly inducing an increased stochasticity in the star-formation episodes \citep{pallottini_MZR_model_2024}: this would dilute the ISM on timescales shorter than required by star-formation to enrich it back to the equilibrium level matched to the average galaxy population at lower redshifts, especially in compact systems \citep{langeroodi_fmr_compactness_2023}. On the other hand, star-formation stochasticity can reflect `internal’ processes associated with SNe-feedback timescales and energetics, which can in turn impact the gas reservoir and the metal content of the ISM \citep{muratov_feedback_FIRE_2015}. Furthermore, the increased merger rate can perturb the gas, driving instabilities which radially mix metals and trigger star-formation \citep{bustamante_merger-induced_2017}.

In conclusion, it is important to note that early JWST studies might be potentially biased by not properly accounting for the complicated selection functions of the different observing programmes carried-out within the first few cycles of observations, convolved with the target allocation constraints set by the Micro-Shutter-Array (MSA) of NIRSpec employed for multi-object spectroscopy, and that confirmation (or disproof) of these initial results shall awaits the analysis of more statistically sound and representative galaxy samples, which will certainly be the subject of multiple studies within the next few years.

\subsection{Abundance patterns of different elements}
\label{ssec:relative_abundances}

The study of relative abundances of different chemical elements, as produced by distinct stellar processes on varying timescales, provides valuable insights into the star-formation history of galaxies. 
In fact, $\alpha$-elements like oxygen (O) and neon (Ne) originate from massive stars dying as core-collapse SNe on timescales of a few Myrs, nitrogen (N) and carbon (C) mainly come from low- to intermediate-mass stars (probing tenths to hundreds of Myrs), while Fe-peak elements
probe much longer timescales (up to $\sim$Gyrs) due to the contribution from Type-Ia SNe \citep{kobayashi_origin_2020}.
Therefore, the ratio of $\alpha$-elements to iron ($\alpha$/Fe) serves as a clock of the star-formation history, in that galaxies showcasing high $\alpha$/Fe ratios must have formed over short timescales, before Type Ia SNe could significantly enrich the ISM, while conversely lower $\alpha$/Fe ratios reflect prolonged star formation periods. 
More in general, abundance patterns of specific elements can reveal the types of stellar progenitors responsible for enriching the ISM, and might even carry the signatures of the enrichment from the first generation of stars, offering insights into the beginning of the cosmic history of metal production.

Detailed studies of chemical abundances in the Milky Way have been performed over the past decades leveraging large, high-resolution spectroscopic surveys in both optical and near-infrared regimes such as APOGEE \citep{Majewski_APOGEE_2017} and GAIA-ESO \citep{Gilmore_GAIA-ESO_2012}, forming the basis of the so-called `Galactic Archaeology' and allowing us to obtain a comprehensive overview of the history of the formation of different Galactic components by studying for instance the distribution of stars over the [$\alpha$/Fe] versus [Fe/H] diagram as a function of both galactocentric distance and height above the Galactic plane \citep{Hayden_alpha_Fe_MW_2015}.

Performing similar studies in extragalactic sources is of course much more complicated. For most galaxies beyond the Local Group, the study of chemical abundances relies on integrated stellar populations; nonetheless, large spectroscopic surveys have provided valuable insights into the chemical composition of nearby galaxies across a wide range of masses and environments.
Early studies based on Lick indices identified an enhancement in $\alpha$/Fe in early-type galaxies compared to stars in the Galactic disk which correlated with stellar velocity dispersion (a proxy for galaxy mass), suggesting that more massive galaxies formed stars earlier and more efficiently over short timescales than their low-mass counterparts (`downsizing' scenario, \citealt{Conroy_gal_arch_2014}); remarkably, this is also been observed at higher redshifts (z$\sim0.7$) from surveys like LEGA-C \citep[][]{Beverage_LEGA-C_2021}.
Cosmological simulations suggest that fast quenching driven by feedback mechanisms (either from SNe or AGN) may effectively shorten star-formation timescales, reproducing the observed $\alpha$/Fe trends in massive systems \citep{Segers_alpha_Fe_EAGLE_2015}. 
Direct measurements of [$\alpha$/Fe] at even higher redshift are extremely challenging, and the high signal-to-noise on the continuum required to derive stellar metallicities is generally reached only in composite spectra. Nonetheless, studies that attempted to perform a self-consistent analysis of rest-frame UV (rich of stellar features) and rest-optical (rich of nebular features) spectra consistently found indications for an $\alpha$-enhancement in $z\sim2-3$ star-forming galaxies up to a factor of $\times~3-4$ compared to the solar value \citep{steidel_reconciling_2016,  Cullen_NIRVANDELS_alpha_enhance_2021}, suggesting these galaxies experienced rapid formation timescales ($\sim0.5–1$~Gyr). Such increase in [$\alpha$/Fe] (linked to a lower metallicity of the stellar populations) was also deemed responsible for the observed offset in the location of $z\sim2-3$ galaxies on several rest-optical diagnostic diagrams compared to local systems, as driven by the hardening of the stellar radiation field ionizing the gas \citep{strom_measuring_2017}. 

\begin{figure}
    \centering
    \includegraphics[width=0.99\linewidth]{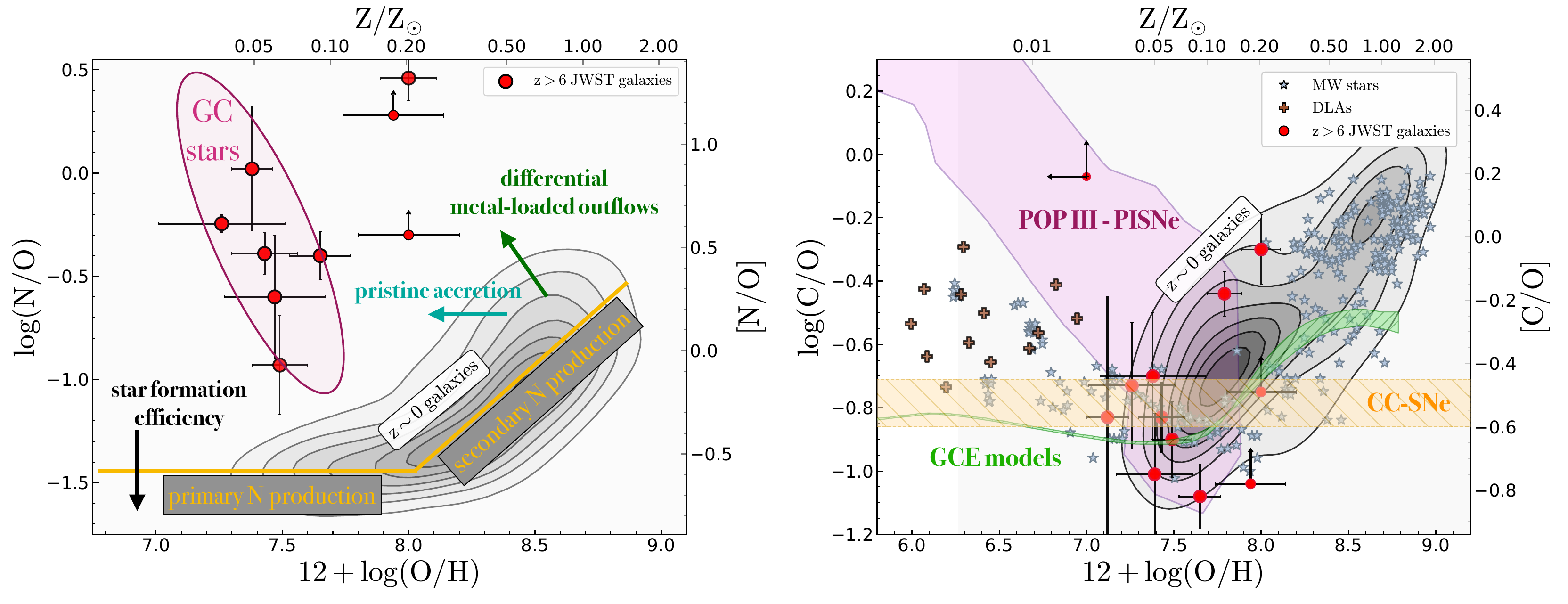}
    \caption{N/O and C/O abundance ratios as a function of oxygen abundance. The left-hand panel reports the distribution of local $z\sim0$ galaxies in the N/O vs O/H diagram, highlighting the primary and secondary production sequence and schematizing some of the physical mechanisms that can possibily affect the relationship. The distribution of N-enriched $z>6$ galaxies observed to date by JWST is also shown: these galaxies showcase super-solar [N/O] and occupy a region of the diagram similar to stars observed within globular clusters in the Milky Way. The right-hand panel reports instead the distribution of Milky Way stars, local galaxies, high-redshift DLAs, and the same JWST sample in the C/O vs O/H diagram. Predictions from pure core-collapse supernovae yields by \citealt{tominaga_2007}, galactic chemical evolution models of the solar neighborhood from \citet{kobayashi_taylor_chemodyn_2023}, and yields from pair-instability supernovae of Pop III stars progenitors from \citet{Vanni_popIII_CO_2023} are also shown.
    }
    \label{fig:abundance_patterns}
\end{figure}

In most systems, especially at high-redshift, the study of relative chemical abundances focuses on the gas-phase, leveraging the presence of bright emission lines of different elements in both optical and UV regimes.
In particular, the relative abundances of nitrogen-to-oxygen (N/O) and carbon-to-oxygen (C/O) are the among the most studied, in virtue of the different nucleosythetic paths of the involved species (Figure~\ref{fig:abundance_patterns}).
For instance, in the low-metallicity regime most of the nitrogen has a primary origin and is generated in massive stars, whereas at higher metallicity nitrogen behave like a secondary\footnote{the yield is proportional to the metallicity of the star at birth, in contrast to primary production in which it is instead independent of the initial metallicity} nucleosynthesis product, since its production becomes dependent on the previous amount of oxygen (and carbon) synthesized in stars via the CNO cycle, hence increasing with increasing oxygen abundance.
From the analysis of large spectroscopic surveys it appears that the local galaxy population is distributed over the N/O vs O/H diagram following a plateau in N/O at low-metallicity (representative of the primary production) and a linear correlation at higher O/H, a clear signature of the contribution from secondary production dominating the nitrogen enrichment \citep{pilyugin_abundance_2012}.
The exact values of plateau, slope of the linear part, and O/H `turnover' depend on the average star-formation efficiency of the analyzed galaxy sample \citep{berg_chaos_2020}, as well as on the methodology adopted to derive the nitrogen abundance (whether this is based on the electron temperature method or on strong-line diagnostics such as $\mathrm{[NII]/[OII]}$ or $\mathrm{[NII]/[SII]}$, \citealt{perez-montero_impact_2009}).
Furthermore, the distribution of galaxies in the N/O vs O/H diagram is also sensitive to gas inflows, which impact the metallicity but not the relative N/O ratio, moving galaxies towards the left, and outflows with differential metal-loading factors (in which oxygen is expelled preferentially by the SN-driven winds compared to nitrogen), which can move galaxies towards the upper-left part of the diagram \citep{vincenzo_nitrogen_2016}. Some of the involved processes and their impact on the distribution of galaxies onto the N/O vs O/H diagram are summarized in the left-hand panel of Figure~\ref{fig:abundance_patterns}.
The N/O has been observed also to tightly correlate directly with stellar mass \citep{masters_tight_2016}, and a possible three-dimensional relationship between $M_\star$, SFR, and N/O (analogue to the FMR derived for oxygen abundance) has been inferred for local galaxies, suggesting that systems that simultaneously obey both the `oxygen-based' and `nitrogen-based' FMR shall experience the same combination of pristine inflows and metal-loaded outflows throughout their evolution \citep{hayden-pawson_NO_2022}. 

Similarly, the C/O vs O/H diagram provides key constraints on the SFH of galaxies, thanks to the delayed release of carbon compared to $\alpha$ elements. Deriving carbon abundances however is challenging, due to the lack of bright optical lines and the need to resort to UV transitions such as $\mathrm{C III]\lambda\lambda1907,09}$ and  $\mathrm{C IV\lambda\lambda1548,51}$, mainly leveraging sensitive spectroscopy with HST instruments like the Cosmic Origins Spectrograph (COS). In order to minimize the uncertainties on the differential reddening correction between optical and UV regimes, the C/O is best derived by comparing the intensity of $\mathrm{C III]\lambda\lambda1907,09}$ to that of the nearby $\mathrm{O III]\lambda\lambda1661,66}$ emission, when also the latter is detected \citep{amorin_analogues_2017}.
In the local Universe, the distribution of galaxies on the C/O vs O/H diagram is characterized by a large dispersion, though with clear similarities to the N/O abundance patterns, showing a low-metallicity plateau while a steep increase at high O/H, mimicking a secondary production process (right-hand panel of Figure~\ref{fig:abundance_patterns}). This similarity is further supported by the fairly constant observed C/N ratio, suggesting a common enrichment origin for carbon and nitrogen \citep{berg_carbon_2016}. Furthermore, the scatter in the C/O vs O/H diagram can be modeled not only by means of variations in the star-formation history (e.g. in terms of number, duration, and delays of bursts) but also including the impact of oxygen-loaded outflows \citep{berg_chemical_2019}.

At high-redshift, despite the inherent observational challenges in detecting relatively faint emission lines such as $\mathrm{[N II]\lambda6584}$ and $\mathrm{C III]\lambda1907,09}$, evolution in N/O and C/O has been studied by multiple spectroscopic surveys conducted with both space- and ground-based facilities, despite these being limited to probing epochs only up to z$\sim2-3$ before the arrival of JWST. 
In the N/O vs O/H diagram, it has been observed that $z\sim2$ galaxies follow trends similar to local galaxies but with larger scatter \citep{strom_nebular_2017, hayden-pawson_NO_2022}, possibly reflecting enhanced star formation efficiencies and inflows of pristine gas. 

Indications on the evolution of C/O abundance at high-z mostly come from deep investigations of Ly$\alpha$ emitters at $z\sim2-4$ with optical facilities like \textsc{VLT/X-SHOOTER} and \textsc{VLT/MUSE} \citep[e.g.][]{vanzella_2016, matthee_2021, citro_CO_z4_2024}, which found an overall consistency in the C/O trend with metallicity as observed in the local Universe.
The most interesting results before JWST followed from studying high-z DLA systems in absorption. In fact, multiple studies reported an enhancement in C/O in DLAs at very low metallicity ($\mathrm{12+log(O/H)<7.5}$) which resembles that observed in some carbon-enhanced metal-poor (CEMP) stars of the Milky Way halo \citep{cooke_dla_2017, saccardi_dla_2023}, and which are considered, according to models,  as some of the best candidate systems polluted by the nucleosynthetic products of so-called Pair Instability Supernovae originating from massive Pop. III stars progenitors \citep{venditti_popIII_2024}.

The systematic investigation of chemical abundance patterns within the epoch of Reionizaion had to await the arrival of JWST. 
Among its most remarkable results so far, JWST unveiled a population of z$\gtrsim 6$ galaxies with extremely bright UV nitrogen lines ($\mathrm{N IV}\lambda1485$, $\mathrm{[N III}\lambda1750$), associated with an enhanced nitrogen enrichment gievn their (relatively low) metallicity \citep[e.g.][]{cameron_gnz11_2023, isobe_CNO_2023, topping_z6_lens_2024}.
Several mechanisms have been proposed to explain such high N/O, including runaway collision of stars in overdense environments \citep{charbonnel_gnz11_2023}, stellar winds from very massive stars (VMS, $\mathrm{M_\star \gtrsim 100 M_\odot}$, \citealt{vink_vms_2024}) or even super massive stars (SMS, $\mathrm{M_\star \gtrsim 1000 M_\odot}$, \citealt{marques-chaves_Nitrogen_2024}), a top-heavy IMF \citep{curti_gs_z9_2024} or stochastic star-formation histories with multiple bursts alternated with lulling periods \citep{kobayashi_ferrara_gnz11_2024}. Overall, the high ISM density and star-formation surface densities observed in such compact, nitrogen-enhanced systems suggest these could be progenitors of present day globular clusters \citep{charbonnel_gnz11_2023, Schaerer_nitrogen_z94_2024}, in virtue of the similar abundance patterns observed in some dwarf stars within local GCs, or even be associated with the formation and presence of early super-massive black holes \citep{maiolino_gnz11_2023, ji_nitrogen_AGN_z5_2024, watanabe_2024}.
Whether such nitrogen enhancement is common at high-redshift or if these observations are only probing systems experiencing a specific (though short-lasting) phase of their evolution will require future, deeper rest-frame UV spectroscopy of the broader (and less luminous) galaxy population. 

In contrast, the vast majority of C/O measurements at $z \gtrsim 6$ delivered by JWST so far report an overall agreement with the distribution of local galaxies at low metallicity, compatible with a recent history of star-formation and a dominant contribution to ISM enrichment driven by core-collapse supernovae. Nonetheless, a notable exception is represented by the galaxy JADES-GS-z12 \citep[][$z=12.5$]{d_eugenio_gsz12_2023}, whose highly enhanced C/O could possibly be modeled as one of the chemical fingerprints of the enrichment from the first generation of PopIII stars.

\subsection{Metallicity gradients in galaxies}

Beyond integrated measurements of the `average' metallicity, the spatial distribution of metals within galaxies offers further, critical insights into their baryonic and chemical assembly, particularly regarding the impact of gas flows and dynamical processes. Tracking the evolution of metallicity gradients over time is therefore essential for understanding galaxy evolution.

In nearby galaxies, metallicity measurements of individual HII regions reveal a dominance of negative\footnote{The metallicity is generally observed to decrease exponentially with galactocentric radius; therefore, its logarithmic form, 12 + log(O/H), varies linearly with radius} metallicity gradients, with central regions more enriched than galaxy outskirts \citep[e.g.][]{bresolin_extragalactic_2009}. 
This supports the so-called `inside-out' mode of galaxy assembly, where compact cores form early through dissipative processes, whereas disks grow at later times. Theoretical models link these gradients to variations in star formation efficiency, gas fraction, and local enrichment dynamics, influenced by inflows and outflows \citep[e.g.][]{carton_gas-phase_2015}. 
At large radii ($\mathrm{R > 2 R_e}$), gradients are often observed to flatten \citep[e.g.][]{sanchez-menguiano_shape_2018}, likely due to dynamical evolution or the recycling of previously ejected metal-rich gas, a phenomenon known as `galactic fountain' effect \citep{fraternali_gas_2017}.
In interacting galaxies, mergers produce flat gradients due to central inflows of metal-poor gas and mixing processes \citep[e.g.][]{kewley_metallicity_2010}.

Large statistical studies performed by integral-field spectroscopic surveys such as CALIFA \citep{sanchez_califa_2012} and MANGA \citep{bundy_overview_2015} suggest that disk galaxies exhibit a characteristic gradient in oxygen abundance when radial distances are normalised to the galaxy effective radius \citep{sanchez_characteristic_2014}. This trend initially appeared largely independent of other galaxy properties, suggesting a common evolutionary pathway.
However, more statistically sound studies (with samples spanning a wider dynamic range in physical parameters) later suggested instead that the metallicity gradients in star-forming galaxies (even when normalized to $R_e$) show a clear dependence on stellar mass \citep{belfiore_sdss_2017, poetrodjojo_sami_2018}: gradients are relatively shallow in low-mass galaxies ($\mathrm{M_\star} \sim 10^{9} M_\odot$), while becoming progressively steeper (more negative) in more massive galaxies. 
However, gradients seem to flatten-out within the central regions of the most massive systems, likely reflecting the saturation in metallicity as predicted also by many analytical GCE models.

At high redshift, the picture becomes less clear, as a consequence of the increased difficulty of performing spatially resolved metallicity measurements. 
Although past studies that achieved high spatial resolution ($\lesssim$~kpc-scale) with the aid of adaptive optics, gravitational lensing, and/or by means of slitless spectroscopy with HST, observed steep negative gradients in a few galaxies at $z\gtrsim 1$ \citep[e.g.][]{jones_keck_2013, wang_grism_2017}, consistent with an `inside-out' growth scenario\footnote{galaxies initially form their central region, producing a steep negative gradient, which then flatten with cosmic time}, larger surveys conducted with ground-based telescopes reported a majority of shallow/flat gradients up to $z\sim2$ \citep[e.g.][]{stott_relationship_2014, carton_inferring_2017, curti_klever_2020}, 
suggesting efficient mixing of gas over large scales possibly driven by enhanced feedback mechanisms or merger-driven radial gas flows.
Nonetheless, the limited spatial resolution pertaining to seeing-limited observations severely hampers the proper determination of radial trends, and beam-smearing effects can play an important role in biasing and artificially flattening the inferred gradients, especially in low-mass galaxies \citep{yuan_systematics_2013, wuyts_evolution_2016}.
Remarkably though, a few examples of `inverted', positive gradients (i.e. with central regions of galaxies more metal depleted than the outskirts) have been observed in high-z galaxies, being interpreted as the signature of cold, pristine gas accretion diluting the ISM in the galaxy core \citep[e.g.][]{cresci_gas_2010, wang_discovery_2019}.

Theoretical frameworks, including analytical GCE models and cosmological hydrodynamical simulations, provide predictions for the evolution of metallicity gradients that can be tested with observations.
Many models consist in the application of the analytical `gas regulator' prescriptions to concentric galactic rings, in which there is a radial dependence of the main parameters such as the timescale of gas infall and the star formation efficiency, whereas other models also invoke possible radial variations in the IMF.
While these models naturally comply with the `inside-out' growth scenario of mass assembly suggested by colour and stellar population age gradients, they predict much steeper gradients at high-z than in the local Universe (at odd with most observations) unless mechanisms that allow gas and metals to be exchanged radially are implemented. 
Possible solutions include stellar migration \citep[e.g.][]{spitoni_stellar_migration_2015}, prominent radial flows of gas that can dilute the central regions at early epochs \citep[e.g.][]{mott_abundance_2013}, or strong feedback that expelled metals produced by central star-formation into the circumgalactic medium and towards the external regions of galactic discs, in line with observations of a relatively enriched CGM at $z\sim2$ \citep[e.g.][]{Prochaska_CGM_2013}.
Similarly, zoom-in cosmological simulations require to include strong feedback mechanisms (including radiation pressure from massive stars) in order to reproduce shallow metallicity gradients at $z\sim2$ \citep[e.g.][]{ gibson_constraining_2013}.
Furthermore, the increased scatter in the observed distribution of gradients at high-redshift possibly reflect the increased stochasticity of the star-formation process at earlier cosmic epochs, coupled with the impact of strong feedback associated with starburst episodes \citep{ma_why_2017}.

Beyond radial trends, azimuthal variations in metallicity (at fixed radius) have been investigated in local spirals, finding evidence for a (small) offset between arm and inter-arm regions \citep{sanchez-menguiano_arm_2017, ho_azimuthal_2018}.
However, it is important to note that the irregular morphologies, multi-component and `clumpy' nature of many high-z galaxies can make the analysis of radial averages in these systems potentially deceiving.
While already suggested in the past by seeing-limited studies \citep{curti_klever_2020}, this is becoming increasingly evident thanks to recent, higher resolution observations performed with the integral-field spectrograph of JWST/NIRSpec, highlighting the importance of studying the full metallicity distribution and its connection to the star-formation properties and kinematics of each galaxy sub-components in order to get a more comprehensive view of galaxy assembly in the early Universe \citep[e.g.][]{del_pino_GANIFS_2024, jones_GANIFS_z57_2024}.

\subsection{Metals in the circumgalactic medium}
\label{ssec:cgm}

Probing the metallicity structure of the circumgalactic medium (CGM) is key to constrain the role of gas flows, cosmic accretion history, and environmental interactions in driving the baryon cycle and its impact onto galaxy evolution.
Because of the much lower density and surface brightness of the gas in the CGM compared to the ISM of galaxies, metallicity measurements via detection of line emission are extremely challenging, despite the progress marked over the past years by deep galaxy surveys with IFU instruments like VLT/MUSE \citep[e.g.][]{dutta_metals_in_haloes_z1_MUSE_2023}.
Therefore, the investigation of the metal content of the CGM mostly relies on absorption line techniques as described in Section~\ref{ssec:absorption_lines}.
Indeed, most of the information on metals in the CGM comes from absorption lines in the UV region of the spectrum from elements like carbon, nitrogen, oxygen, silicon, iron, magnesium, calcium.

The CGM is a multi-phase medium, constituted by a wide range of density, temperature, and ionization conditions. 
Simulations reveal structured CGM gas at high redshifts, with accreting gas in filaments and outflows forming biconical patterns \citep{corlies_metals_cgm_2016}. 
However, at lower redshifts, the CGM becomes more mixed, and gas velocities provide less insight into its origin \citep{Christensen_gas_flows_2016}.
In general, absorption spectroscopy using the galaxy’s own starlight as a background source (so-called `down-the-barrel' technique) is capable to distinguish inflows from outflows by interpreting blueshifted absorption as tracing outflowing gas and redshifted absorption as tracing inflowing gas, respectively; it mainly probes gas at the disk-halo interface, but the galactocentric radius of any detected absorption is basically unconstrained.
On the other hand, transverse absorption spectroscopy using distant quasars provides a more direct measure of the distance between the galaxy and the probed sightline, but the direction of the gas flow is degenerate such that blueshifted absorption might be related to infalling gas behind the galaxy or equally to outflowing gas in front of the galaxy, and viceversa for redshifted absorption.
Combining both observations (when possible) can increase the complexity of the interpretation but also provide deeper insights into the dynamics of the CGM than either alone.

The metal content of galactic flows can serve in principle to distinguish their origin, in that one would expect infalling gas to be nearly-pristine while outflows to be metal-enriched due to star-formation processes taking place inside galaxies \citep[as observationally suggested in the past by the study of a large sample of LLSs revealing a bi-modal metallicity distribution, e.g.][]{Lehner_bimodal_metallicity_CGM_2013}, or to distinguish between systems associated with either the CGM of galaxies or with the IGM \citep{BergM_BASIC_2023}.
Gas accreting onto galaxies from the cosmic web passes through the CGM on its journey from the IGM to galaxies, where it leaves observable signatures that can be exploited to trace the inflows \citep{Tumlinson_cgm_review_2017}.
In this sense, cold, dense, metal-poor CGM gas observed in the form of LLS is considered a strong indicator of cosmological gas accretion at all redshifts, as also predicted by simulations \citep{Hafen_LLS_inflows}.

On the other hand, we have seen that galactic outflows represent a critical component for most galaxy evolution models.
Indeed, large-scale outflows have been observed at almost all cosmic epochs \citep{rupke_outflows_review_2018}, providing feedback that influences the star-formation process in galaxies and carrying metals far from their formation site, possibly into the circumgalactic medium or even beyond \citep{peroux_howk_CGM_2020}.
For instance, observations of widespread $\mathrm{O VI}$ around galaxies as detected by COS-Halos surveys confirmed the presence of extended metal-enriched outflows up to $\sim300$~kpc \citep{Johnson_quasar_host_haloes_2015}, and simulations predict that the time-integrated effect of enriched outflows are necessary to produce the observed reservoir of metals \citep{suresh_cgm_metals_2017}.
Probing the metal enrichment in the CGM is also relevant to constrain the nature of the mechanisms powering outflows in galaxies, discriminating between `momentum-driven' (with velocity scaling as $\mathrm{v \propto v_{circ}^{-1}}$) and `energy-driven' ($\mathrm{v \propto v_{circ}^{-2}}$) winds, which can produce for instance different slopes of the mass-metallicity relation \citep{finlator_dave_2008, curti_jades_mzr_2024}.

Furthermore, it is possible that a substantial fraction of gas accreted onto galaxies has been recycled from previously ejected (and enriched) material, as predicted by several simulations \citep[e.g.][]{Muratov_metal_budget_2017}, although the timescales of the process and their dependence on the halo mass are not fully constrained.
Recycled accretion might arise from either the ejection of metal-enriched galactic winds lacking the energy to escape the halo, 
from the same winds losing energy by shocks, mixing into the CGM and then eventually cooling and raining back onto the galaxy, or from pristine inflows entraining metal-polluted CGM material on its way from the IGM to the galaxy \citep[e.g.][]{fraternali_gas_2017, Tumlinson_cgm_review_2017}.
Despite being often neglected in the framework of `gas-regulator' chemical evolution models, metal-enriched recycling might therefore constitute an important mechanisms for galaxies to acquire their gas.

\subsection{The metal budget}
\label{ssec:metal_budget}

Beside the metallicity (i.e. a relative measurement of the abundance of heavy element compared to hydrogen), accounting for the total mass of metals produced by stars in galaxies over their lifetime and their relative distribution within different components (stars, ISM, dust, CGM, etc..) provides a useful benchmark to test our understanding of the cosmic star-formation history and the role of gas flows in galaxy evolution.
These are no less than difficult measurements, as inferring the total content of metals often implies having a good knowledge of the total mass associated with all the involved phases. 

\begin{figure}
    \centering
    \includegraphics[width=0.99\linewidth]{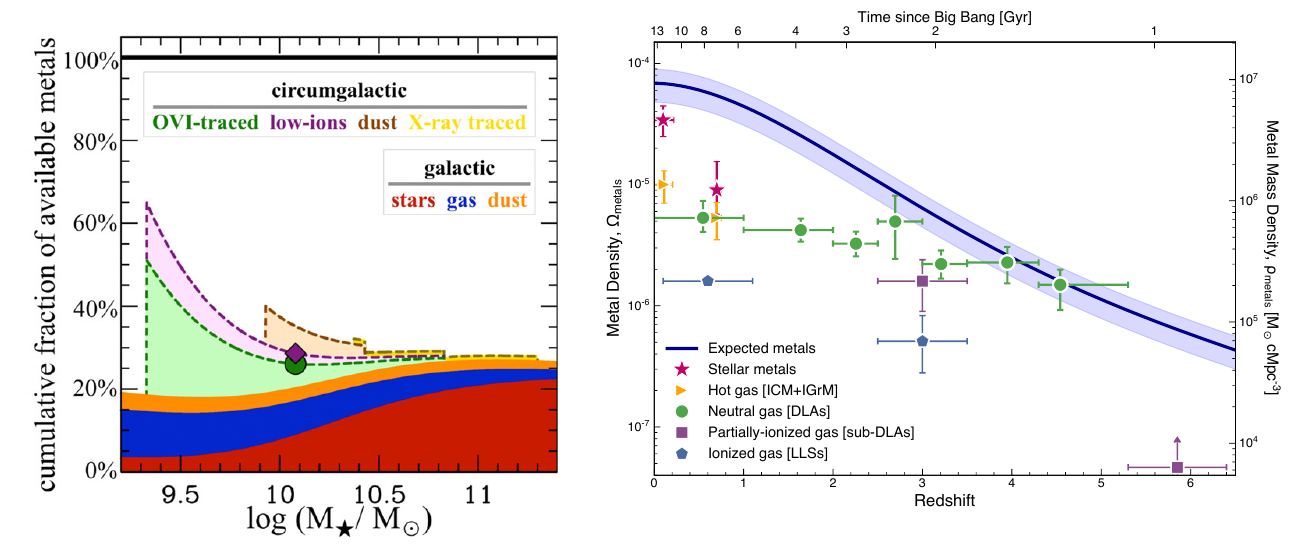}
    \caption{A census of metals observed in different phases. The left-hand panel reports the cumulative fractions of available metals as a function of stellar mass at z$=0$, whereas the right-hand panel traces the evolution of cosmic metal densities in different components as a function of redshift. Reproduced with permission from \citealt{Peeples_metal_budget_2014} and \citealt{peroux_howk_CGM_2020}.}
    \label{fig:metal_budget}
\end{figure}

Historically, large efforts in this area were triggered by early indications for the observed amount of metals in stars and neutral gas to be significantly lower (about one order of magnitude) than expected based on the integrated star formation rate density evolution of the Universe \citep{pettini_chemical_evolution_1999}. 
This `missing metals problem' has been then re-assessed multiple times ever since, leveraging the increased number of observational constraints spanning a wide range of time in cosmic history.
For instance, a full census of metals in galaxies at z$\sim0$ was performed by \cite{Peeples_metal_budget_2014}, which found that the amount of metals found in stars, interstellar gas, and interstellar dust (i.e. those retained inside galaxies) represent only $20–30\%$ of the total metal production over three orders of magnitude in stellar mass (modulo the uncertainties on the assumed stellar yields), indicating that galaxies lose the majority of heavy elements they produce (Fig.~\ref{fig:metal_budget}, left panel)
Including metals observed in the different phases of the CGM (out to $\sim150$~kpc\footnote{hence sampling different fractions of the virial radius with respect to stellar mass}) it is possible to account for about half of the metals produced by typical star-forming galaxies.
Most of the metals in the CGM are in a highly-ionized phase, traced e.g. by O IV \citep{Tumlinson_metals_CGM_2011}, or depleted onto dust, whereas the relative contribution of the hot-CGM (T$>6\times10^6$~K) gas component is small compared to the ISM phase.

Furthermore, by combining absorption-line derived metal abundances (accounting for dust depletion and ionization corrections) with the cosmological density of the same absorbing systems, it is possible to self-consistently assess the contribution of neutral and ionized gas to the total metal content of the Universe as a function of redshift, and compare it to the total comoving metal mass density produced by stars at each given cosmic epoch.
Such a revisited, detailed census has been presented by \cite{peroux_howk_CGM_2020} (see also \citealt{yates_metal_budget_2021} for a comparison with predictions from different cosmological simulations), which found that metals in the neutral gas phase dominate over the ionized component and show a mild evolution with redshift (Fig.~\ref{fig:metal_budget}, right panel).
At the higher redshifts ($\mathrm{2.5<z<5}$), the vast majority all of the expected metals are found in the neutral phase (with a relatively significant contribution of ionized absorbers at z$\sim3$), whereas at lower redshifts (z$\lesssim 1$) the contributions are more diverse, with the stars dominating the metal budget but with also significant contribution from the hot gas within the intracluster medium (ICM) and intragroup medium (IGrM). 
Although there are still specific redshift ranges that needs to be properly characterized due to intrinsic observational challenges in probing metals in their different phases (and which shall await future X-ray/UV facilities such as \textsc{ATHENA} and \textsc{LUVOIR}, especially for the hot gas component), the `missing metals problem' is therefore largely reduced as of today, and the expected metal content of the Universe is now largely accounted for, once the large uncertainties ($\approx 30\%$) on the assumed stellar yields are considered.

\section{Conclusions and future prospects}
 
In this chapter, we have provided an overview of the key methodologies, observational findings, and their constraints on models that form the foundation of the study of chemical evolution in galaxies across cosmic time. Over the past decade, advancements in ground- and space-based facilities, culminating in the launch of JWST in 2021, have addressed long-standing challenges in the field. These efforts have linked the chemical enrichment properties of galaxies across different spatial and temporal scales, while also delivering surprising results that challenge prior interpretations, offering novel and more robust constraints on models and pushing the boundaries of our understanding.

The future of this field is exceptionally promising, with major progress expected on several interconnected fronts in the coming years.
Over the next decade in fact, the investigation of the metal content in galaxies will proceed over parallel, though deeply interconnected, avenues.
On the one hand, JWST will continue its groundbreaking exploration of the early universe, unveiling new peculiar systems, constraining enrichment mechanisms, and probing the timescales of chemical evolution in the earliest galaxies. Its expanding datasets will enable more robust statistical study of metallicity scaling relations at high redshift, establishing robust connections between the chemical properties of galaxy populations at different cosmic epochs.

On the other hand, next generations, high-multiplexing, optical multi-object spectrographs on 4-meter class telescopes, such as 4MOST on ESO/VISTA, will deliver millions of high-resolution spectra, spanning $\sim 7$~billion years of cosmic history. These observations will tightly link the chemical properties of galaxies to their location within the cosmic web, providing unprecedented insights into the relationship between environment and metal enrichment. Similarly, near-infrared spectrographs like PFS on Subaru and MOONS on the VLT will extend these surveys to higher redshifts ($\mathrm{z \sim 1-2}$), replicating SDSS-like studies at `Cosmic Noon'\footnote{the epoch of the peak of the cosmic star-formation rate density, \citet{madau_cosmic_2014}} while marking an unprecedented improvement in statistics and completeness at these epochs. These facilities will allow for detailed examinations of environmental effects on metallicity scaling relations (for both gas-phase and stellar metallicities), as well as on quenching mechanisms and their associated timescales by linking the metal content of local passive systems with that of their `direct', high-z star-forming progenitors \citep[e.g.][]{peng_strangulation_2015, Trussler_starvation_2020}.

Finally, within the next decade, the advent of extremely large telescopes (ELTs) will mark a transformative step in galactic chemical evolution studies. The ESO E-ELT, featuring a 39-meter primary mirror, will be the first of its kind and promises to arxrevolutionize multiple areas of research with its unparalleled light-collecting power, cutting-edge instrumentation, and deeply integrated support from adaptive optics.
For instance, MICADO will provide both imaging and single-slit NIR spectroscopy, delivering spectra of individual HII regions in $z\sim1-2$ galaxies, while the optical/NIR IFS HARMONI will directly map stellar populations, abundances, and kinematics at $100$~pc resolution in $z\sim2$ galaxies, capturing the most detailed, spatially resolved view of the epoch of the peak of galaxy assembly. 
Leveraging the huge collecting area of the telescope, MOSAIC will probe metals in the CGM at high-z from absorption spectra using relatively faint galaxies (rather than rare, bright quasars) as background sources, hence increasing the number of observable systems at high spectral resolution by orders of magnitude, while the high-resolution spectrograph ANDES ($R\sim100,000$) will search for direct signatures (i.e. Hydrogen lines with no metal transitions) of Population III stars, shedding light on the nature of the systems responsible for the synthesis of the first metals in the history of the Universe.

\bibliographystyle{Harvard}
\bibliography{els-article}

\end{document}